\documentclass{article}

\usepackage{amsfonts,amssymb,graphicx,amsmath}

\baselineskip
\advance \textheight by \topskip
\makeatletter
\@addtoreset{equation}{section}

\makeatother

\begin{document}

\title{\textbf{Potts model on recursive lattices: some new exact results}}
\author{Pedro D. Alvarez$^{1}$\thanks{alvarez AT cecs.cl}, Fabrizio Canfora$^{1}$\thanks{canfora AT cecs.cl}, Sebasti\'{a}n A. Reyes$^{2}$\thanks{sreyes AT fis.puc.cl}, Simon Riquelme$^{2}$ \\
\\
{\small $^{1}$\textit{Centro de Estudios Cient\'{\i}ficos} (\textit{CECs), Valdivia, Chile}} \\
{\small $^{2}$\textit{Facultad de F\'{\i}sica, Pontificia Universidad Cat\'{o}lica de Chile, Casilla 306, Santiago 22, Chile}}}

\maketitle

\begin{abstract}
We compute the partition function of the Potts model with arbitrary values of $q$ and temperature on some strip lattices. We consider strips of width $L_y=2$, for three different lattices: square, diced and `shortest-path' (to be defined in the text). We also get the exact solution for strips of the Kagome lattice for widths $L_y=2,3,4,5$. As further examples we consider two lattices with different type of regular symmetry: a strip with alternating layers of width $L_y=3$ and $L_y=m+2$, and a strip with variable width. Finally we make some remarks on the Fisher zeros for the Kagome lattice and their large $q$-limit.\\

Keyword: Potts model, exact results, kagome lattice, diced lattice, dichromatic polynomial.

PACS: 12.40.Nn,11.55.Jy, 05.20.-y, 05.70.Fh.
\end{abstract}

\section{Introduction}\label{intro}

The Potts model \cite{Potts_52} is one of the most interesting models in statistical mechanics. It appears as the most natural generalization of the Ising model by allowing the spin variable to take more than two distinct values. Besides its intrinsic importance in the theory of critical phenomena, it manifests a number of intriguing relations with many areas of both physics and mathematics (for nice detailed reviews see \cite{Wu82,Baxter,RevModPhys.64.1099,Yang_94,Cardy01}; for the relation with the problem of color confinement see \cite{Svetitsky1982423}; for its relation with the Khovanov Homology see \cite{2009arXiv0907.3178K}).

We will study the $q-$state Potts model at zero magnetic field, whose partition function on an arbitrary graph $G$ is,
\begin{equation*}
Z(G,q,\beta )=\sum_{\left\{ \sigma _{i}\right\} }e^{-\beta \mathcal{H}}
\end{equation*}
where $\beta =1/k_{B}T$ and
\begin{equation*}
\mathcal{H}=-J\sum_{\left\langle ij\right\rangle }\delta _{\sigma _{i}\sigma
_{j}}.
\end{equation*}
Spin variables $\sigma _{i}$ can take values $1,...,q$ and are located at the vertices of $G$. There is also an interaction energy $J$ for pairs of spins located at the ends of every edge $\left\langle ij\right\rangle \in G$. It will be important for our development to introduce an equivalent expression for the partition function,
\begin{equation}\label{p_func_2}
Z(G,q,v)=\sum_{\left\{ \sigma _{i}\right\} }\prod_{\left\langle
ij\right\rangle }\left( 1+v\delta _{\sigma _{i}\sigma _{j}}\right),
\end{equation}
where the temperature variable $v=e^K-1$ has been introduced with $K\equiv \beta J$. For the ferromagnetic case ($J>0$) one has $0\leq v\leq \infty $ corresponding to the temperature interval $\infty \geq T\geq 0$; for the antiferromagnetic Potts ($J<0$) the interval $-1\leq v\leq 0$ corresponds to $0\leq T\leq \infty $. It can be shown directly from (\ref{p_func_2}) that the partition function admits also the following polynomial (Fortuin-Kasteleyn) representation \cite{Fort_Kast_72,Fort_Kast_69},
\begin{equation}\label{F-K}
Z(G,q,v)=\sum_{G^{\prime }\subseteq G}q^{k(G^{\prime })}v^{e(G^{\prime })}
\end{equation}
where $G^{\prime}$ is a graph with the same vertex set $V$ of $G$ and an edge set $E^{\prime}\subseteq E$, with $E$ being the edge set of $G$; $k(G^{\prime})$ and $e(G^{\prime})$ are respectively the number of connected components and the number of vertices of $G^{\prime }$. It is important to notice that using (\ref{F-K}) one can extend $q$ and $v$ from their physical values ($q\in \mathbb{Z}_{+}$, $v\in \lbrack -1,+\infty )$) to the whole complex plane.

Unlike the case of the Ising model ($q=2$), solved by Onsager \cite{PhysRev.65.117}, the exact solution for the Potts model in any infinite two dimensional lattice is still not available. Nevertheless, thanks to universality, the critical behavior of the ferromagnetic Potts model is fairly well understood. On the other hand, the antiferromagnetic Potts depends on the structural properties of the lattice and its critical properties can vary strongly from case to case. Thus, the search for tools to obtain analytical information on the free energy of the Potts model on generic lattices is a very important task.

An approach that has proven fruitful in recent years is to solve the simpler problem of calculating the partition function for periodic strips of finite width ($L_y$) and arbitrary length ($L_x$). Previous work along these lines has relied basically on a transfer matrix method \cite{BLOTE82,Shrock2000388,Salas01,Salas_01_2,Chang02} to obtain exact results at arbitrary temperatures for strips of widths going up to $L_y=7$ (honeycomb lattice with free boundary conditions) \cite{Chang08}. Such efforts have produced detailed studies for the square, triangular, and honeycomb lattices \cite{Chang2001234,Salas01,Salas_01_2,Chang02,Chang04,Chang2001183,Chang08,Salas09}.

In many cases it is useful to exploit the recursive structure which is present in lattices of physical interest. Even when the analytic solution is not available, using a simple ansatz which respects the recursive symmetry one can get an excellent agreement with the numerical data \cite{Canfora200754,Astorino2008139,Astorino2009291,Canfora_10}. Motivated by these facts we develop a method that permits the calculation of the exact partition function for strips of layered lattices.

The idea to use the dichromatic polynomial in the cases of recursive lattices has been proposed in \cite{Biggs_72,Beraha_79,Beraha_80}, where the authors derive a scalar homogeneous recurrence equation of order higher than one, whose solution is the dichromatic polynomial of the corresponding graph. In the present paper it will be shown that it is possible to extend this idea by formulating vectorial homogeneous order one recurrence equations whose solutions represent the dichromatic polynomials of graphs corresponding to different boundary conditions. The matrix associated to this linear system of equations is equivalent to the transfer matrix in the basis of connectivities at the top layer used by Salas and Sokal (cf. \cite{Salas01}).

Lastly, it is worth emphasizing from the outset that the algorithm developed below was implemented in a computer software in order to handle wider strips or more complicated lattices. Furthermore, it will be shown in the following sections that our formulation also allows to deal with some recursive lattices which are regular, but not necessarily translation invariant \cite{bedini}.

The paper is organized as follows: In Sec. \ref{method} we explain the general method and reproduce some known results \cite{Shrock2000388,Chang02}. Next, in Sec. \ref{simple} we illustrate the use of the method by solving a few different narrow strips of width $L_y=2$. We present new exact results for arbitrarily long strips of width $L_y=2$ for the kagome, diced and `shortest path'\footnote{This lattice will be introduced in Sec. \ref{sh_path}. It corresponds to a square lattice plus a graph inside each square that has a topology identical to that of the shortest path joining all four corners.} lattices. In order to demonstrate the versatility of the procedure, in Sec. \ref{ornamented} we obtain the exact solution of a $L_y=3$ strip corresponding to a mixture of triangular and square lattices, which additionally has periodic attachments of segments of a different strip. Lastly, in Sec. \ref{step-dependent} we analyze an interesting case of a non-periodic lattice for which the transfer matrix becomes step-dependent. Conclusions and prospects for future work are discussed in Sec. \ref{conc}.

\section{Recursive equation method}\label{method}

The Potts model partition function on a graph $G$ ($Z_G$) satisfies the well known deletion-contraction theorem, which says that it can be written in terms of $Z_{G_d}$ and $Z_{G_c}$ as,
\begin{equation}\label{del_cont}
Z_G=Z_{G_d}+vZ_{G_c},
\end{equation}
where $G_d$ corresponds to the resulting graph when deleting an arbitrarily chosen edge from $G$; and $G_c$ is obtained from $G$ by contracting the same edge (i.e. deleting the edge and identifying the vertices at the end of it).
Furthermore, it is easy to show that the Potts model partition function also fulfills the property $Z_{(\bullet \ G)}=q Z_{G}$, where $Z_{(\bullet \ G)}$ denotes the partition function of a disjoint union of an isolated vertex and a graph $G$. The resulting partition function is a polynomial in $q$ and $v$ which in the the mathematical literature is known as the dichromatic polynomial of $G$.

\begin{figure}[t]
\begin{center}
\includegraphics[width=0.2\textwidth]{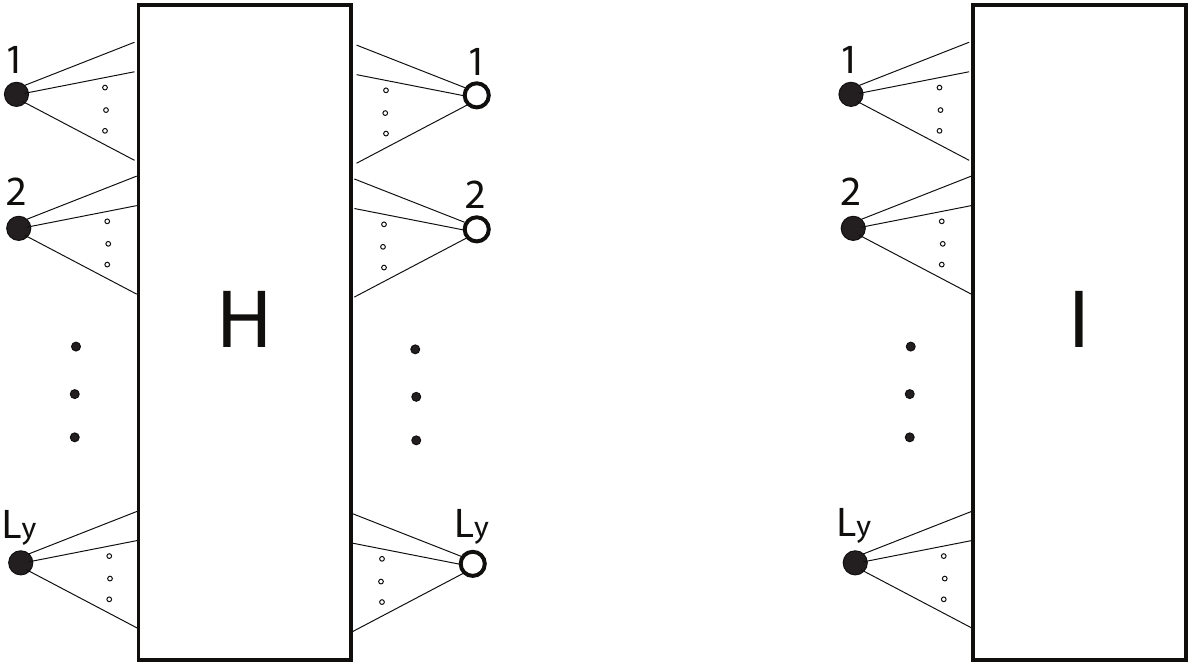}\\[0pt]
\hspace{0.4cm}(a)\hspace{2cm}(b)\\[0pt]
\includegraphics[width=0.47\textwidth]{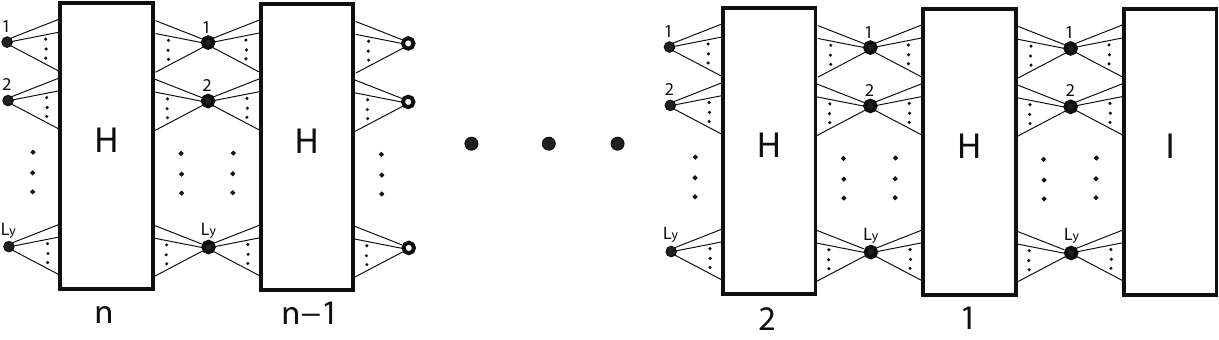}\\[0pt]
(c)\\[0pt]
\end{center}
\caption{Basic building blocks for a periodic lattice strip. (a) $H$ corresponds to the structure that will be repeated $n$ times. Empty circles to the right denote vertices that will be shared between adjacent blocks. (b) Graph $I$ at the right end of the strip containing the information of the initial condition $\vec{Z}(0)=\vec{Z}_{I}$. (c) The strip is built up by repeatedly joining $H$-blocks after $I$.}
\label{periodic_G}
\end{figure}

Using the notation of \cite{ROEK98,Rocek98}, consider now strip graphs of the form $(G_{s})_{n}=(\prod_{l=1}^{n}H)I$, where $H$ is repeatedly attached after the initial subgraph $I$, which will be put to the right by convention. The junction between contiguous subgraphs is done by sharing a subset of $L_y$ vertices ($\tilde{V}$) as illustrated in Fig. \ref{periodic_G}.
Each block $I$ or $H$ can contain an arbitrary number of vertices besides the ones in $\tilde{V}$. Suppose now that starting from a strip of length $n$ as the one shown in Fig. \ref{periodic_G}c, we apply repeatedly the deletion-contraction theorem to all the edges in the last $H$ block. Once this process is finished, it is easy to see that the partition function of the initial graph ($Z_{0}(n)$) will be written as a linear combination of the partition functions of objects of length $n-1$,
\begin{equation*}
Z_{0}(n)=\sum_{j=0}^{N-1}a_{0j}(q,v)Z_{j}(n-1) \,,
\end{equation*}
where the coefficients $a_{0j}(q,v)$ are polynomials in $q$ and $v$. The number of coefficients $N$ and their expression is deduced only by the  application of the deletion-contraction on the $n$-th block $H$, no extra information is needed. Clearly $N$ counts as well how many configurations $Z_j(n)$ arise from the different ways in which the $L_y$ vertices in $\tilde{V}$ between the $n$-th and $(n-1)$-th $H$-blocks become identified with each other, after the action of the deletion-contraction over the whole $n$-th block. Then one should proceed analogously starting with each one of the possible $Z_{j}(n)$ until the system is closed. The emerging linear system of equations can be written as
\begin{equation}\label{linear_syst}
\vec{Z}(n)=\mathbf{A}(q,v)\vec{Z}(n-1)\,,
\end{equation}
where $\mathbf{A}(q,v)$ is the $N \times N$ transfer matrix with elements $a_{ij}(q,v)$ and the vector $\vec{Z}(n)$ contains the partition functions for strips of length $n$ with all possible different connectivities at the top layer. It is worth noting that $\mathbf{A}(q,v)$ is independent of the step $n$ whenever the block $H$ is the same for all layers. As it will be explained in detail below, the dimension $N$ of $\mathbf{A}(q,v)$ can be determined considering only some basic properties of $H$ (i.e. symmetries or if it is planar or not). By recursive use of (\ref{linear_syst}), it is straightforward to show that the general solution is
\begin{equation}\label{linear_syst2}
\vec{Z}(n)=\mathbf{A}^{n}(q,v)\vec{Z}(0),
\end{equation}
where $\vec{Z}(0)$ is the partition function vector corresponding to the step$-0$ block $I$. In the context of the recursive equations (\ref{linear_syst}) $\vec{Z}(0)$ plays the role of initial conditions. Remember that the sought partition function $Z_0(n)$ is simply the first component of the vector $\vec{Z}(n)$. A convenient expression for it is
\begin{equation}\label{Z(n)}
Z_0(n)=\sum_{j=1}^{N}\alpha_j(q,v)\lambda _{j}^{n}(q,v)\,,
\end{equation}
where $\lambda _{j}$ are the eigenvalues of $\mathbf{A}$ and the $\alpha_j$ coefficients are sensible to the lattice structure and to the initial vector $\vec{Z}(0)=\vec{Z}_I$. 
Expression (\ref{Z(n)}) is useful to study the thermodynamic limit $n\rightarrow \infty$, in which the relevant free energy should be normalized by the total number of vertices of the lattice. An exact expression for the coefficients in terms of the eigenvalues and the lattices of length $1,2,...,n$ can be obtained,
\begin{equation}\label{coefs}
\alpha_i(q,v)=\frac{1}{\det{\mathcal{M}}}\sum_{j=1}^{N}\mathcal{M}^*_{ji}Z_0(j) \,,
\end{equation}
where $\mathcal{M}^*_{ij}$ is the $(i,j)$-th cofactor of a matrix
\begin{equation}
\mathcal{M}=\left(
\begin{array}{llll}
1 & 1 & \ldots & 1\\
\lambda_1 & \lambda_2 & \ldots & \lambda_N\\
\vdots & \vdots & \ddots & \vdots\\
\lambda^{N-1}_1 & \lambda^{N-1}_2 & \ldots & \lambda^{N-1}_N
\end{array}
\right) \,.
\end{equation}
The behavior of $Z_0(n)$ in the thermodynamic limit is given by the dominant eigenvalue associated to a nonzero $\alpha_j$ coefficient. When there are no degeneracies we will denote it by $\lambda_+$, and the sub-dominant by $\lambda_-$. Clearly, when none of the coefficients given by (\ref{coefs}) are zero we have  $|\lambda_+|\ge|\lambda_-|\ge|\lambda_j|$, $j \in \{ 1,...,N \}\ne \{\pm\}$.

When one deals with a family of graphs $\mathcal{G}=\{G_1,G_2,...\}$ is sometimes useful to work with a generating function,
\begin{equation}
 \Gamma(\mathcal{G},q,v;y)=\sum_{n=1}^\infty y^{-n} Z(n-1) \,,
\end{equation}
which is a formal expansion in terms of an auxiliary variable $y$. This approach has been proposed and applied for zero temperature calculations in \cite{ROEK98,Shrock1998315,Rocek98}. The generating functions for the set of all possible identifications at the top layer can be easily computed from the transfer matrix $\mathbf{A}(q,v)$ as follows
\begin{equation}
 \vec{\Gamma}(\mathcal{G},q,v;y)=(y-\mathbf{A}(q,v))^{-1} \vec{Z}(0) \,.
\end{equation}
Lastly, it is useful consider the following expression for the generating function,
\begin{eqnarray}\label{Gamma}
\Gamma(\mathcal{G},q,v;y)&=&\frac{1}{\det[ y-\mathbf{A}]}\sum_{j=1} ^N (y-\mathbf{A})_{j,1}^* Z_{j-1}(0)\nonumber\\
&\equiv&\frac{\mathcal{N}(\mathcal{G},q,v;y)}{\mathcal{D}(\mathcal{G},q,v;y)} \,.
\end{eqnarray}
We observe that $\Gamma(\mathcal{G},q,v;y)$ is a rational function and that its denominator $\mathcal{D}(\mathcal{G},q,v;y)$ has degree $N$ in $y$. Furthermore, from (\ref{Gamma}) it is clear that the denominator is the characteristic polynomial for the eigenvalues of $\mathbf{A}(q,v)$, i.e. $\mathcal{D}(\mathcal{G},q,v;y)=\Pi_{j=1}^N (y-\lambda_j)$. The numerator $\mathcal{N}(\mathcal{G},q,v;y)$ turns out to be a linear function of $y$ where corresponding coefficients depends both on the structure of the block $H$ and the boundary configurations $Z_j(0)$.

\section{Simple applications}\label{simple}
For illustration purposes, in this section we apply the method to solve a few simple examples.

\subsection{$L_y=2$ square ladder}

\begin{figure}[t]
\begin{center}
\includegraphics[width=0.10\textwidth]{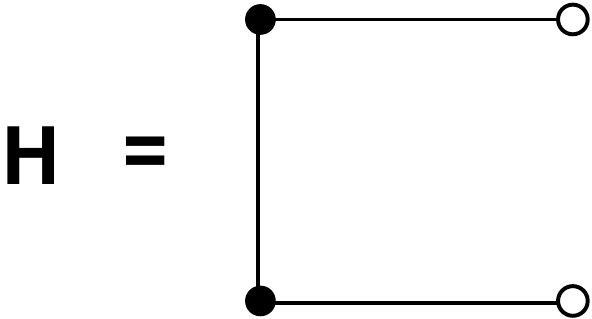}~~~~~~~~~~~~
\includegraphics[width=0.05\textwidth]{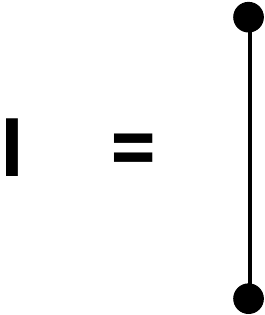}\\[0pt]
\hspace{1.2cm}(a)\hspace{2.3cm}(b) \\[0pt]
\vspace{0.4cm} \includegraphics[width=0.3\textwidth]{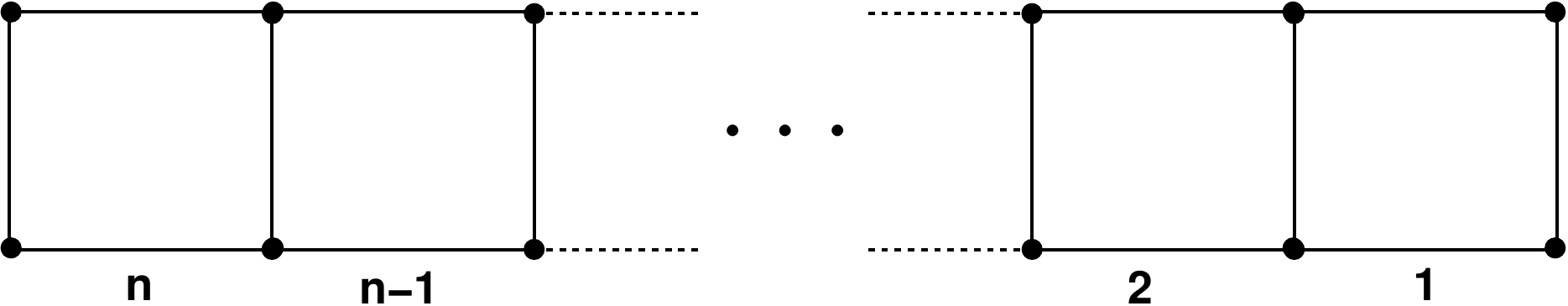}\\[0pt]
(c)
\end{center}
\caption{For the square ladder it is quite simple to find its basic building blocks. (a) Every $H$ adds two new vertices (full circles) and three edges. A block is joined to the previous one by sharing the $L_y=2$ vertices to the left (empty circles). (b) In this case the initial graph $I$ consists of two joined vertices. }
\label{ladder}
\end{figure}

\begin{figure}[t]
\begin{center}
\includegraphics[width=0.2\textwidth]{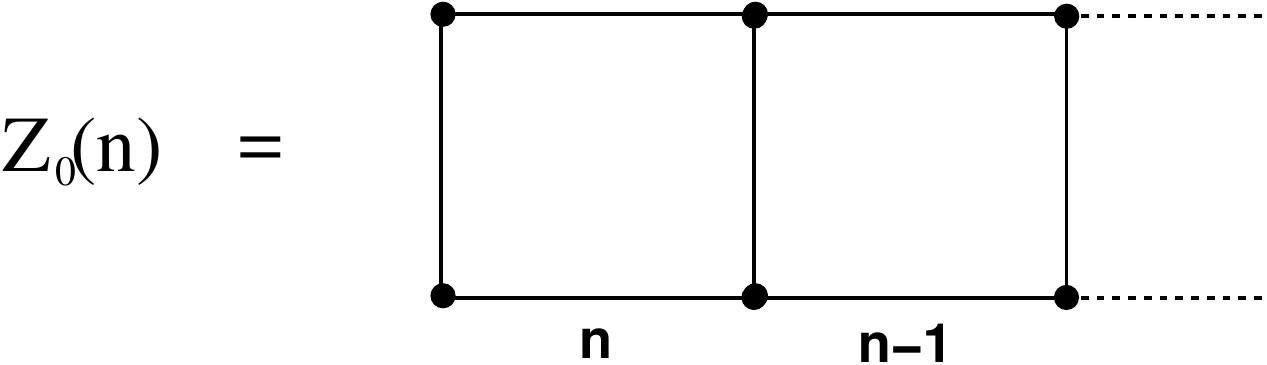}~~~~~~~~~~~~
\includegraphics[width=0.2\textwidth]{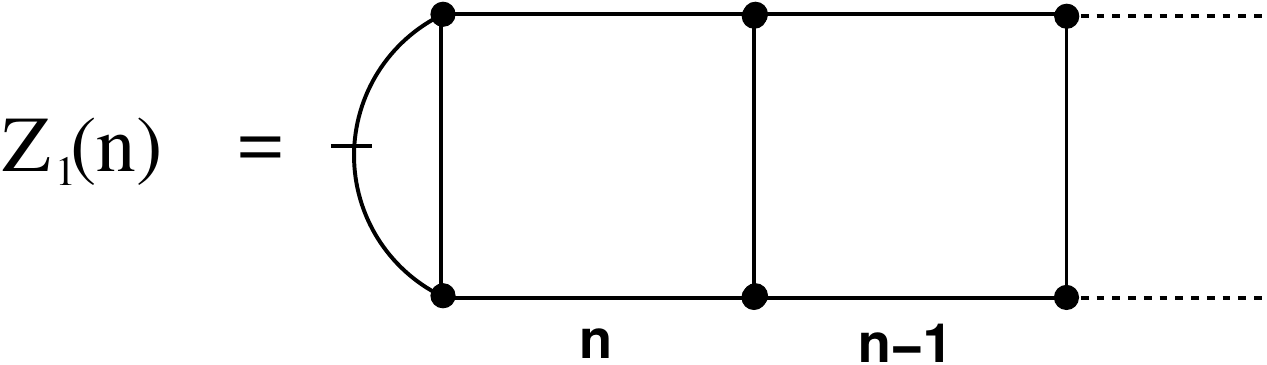}\\[0pt]
\hspace{1cm}(a)\hspace{4cm}(b) \\[0pt]
\end{center}
\caption{(a) $Z_0(n)$ is the sought partition function which in our convention is the first component of $\vec{Z}(n)$. None of the end vertices of the graph are identified with each other. (b) Once the deletion-contraction theorem is applied to the previous block, a new kind of graph will appear in which the two leftmost vertices are identified with each other. In our notation identified vertices are joined by a crossed line.}
\label{basis}
\end{figure}

Here we will consider the case of $2\times n$ square ladder lattices.\footnote{We solve this system for illustration purposes only since the solution was already obtained in \cite{Shrock2000388}.}
As illustrated in Fig. \ref{ladder}, it is easy to find for this lattice its corresponding initial graph $I$ and its building block $H$.\footnote{It is important to observe that there is a certain arbitrariness in the choice of $H$. For example, one could choose instead to join the graphs by the two vertices along a diagonal of the square.}
For this case we have $L_y=2$ and by direct calculation it is easy to see that $N=2$ as explained in Fig. \ref{basis}. Here we only have the original graph corresponding to $Z_0$ and the graph with the two vertices at the end identified with each other. The vector of the initial graph $I$ is then,
\begin{equation}\label{Z0_sq_2}
\vec{Z}(0)= \left[
\begin{array}{c}
q(q+v) \\
q(1+v)
\end{array}
\right].
\end{equation}

In order to construct matrix $\mathbf{A}^{\text{sq}}(q,v)$ for the square (sq) strip, we need to delete and contract all three edges in $H$ and repeat the procedure starting from $Z_{1}(n)$ to obtain finally,
\begin{equation*}
\mathbf{A}^{\text{sq}}(q,v)=\left[
\begin{array}{cc}
(q+v)(q+2v)+v^{2} & v^{3} \\
(1+v)(q+2v) & (1+v)v^{2}
\end{array}
\right].
\end{equation*}

It is now trivial to diagonalize this $2\times 2$ matrix to find that its eigenvalues are
\begin{equation}
\lambda_{\pm }^{\text{sq}}(q,v)=\frac{1}{2}(q^{2}+3qv+4v^{2}+v^{3}\pm\sqrt{q^{4}+6q^{3}v+13q^{2}v^{2}+16qv^{3}-2q^{2}v^{3}+12v^{4}-2qv^{4}+4v^{5}+v^{6}}).
\end{equation}

As it is well known, the dominant eigenvalue $\lambda_+$ will determine the free energy in the thermodynamic limit. It can be noted that the discriminant is always positive in the physical regime.

\subsection{Diced lattice strip}\label{diced}

The diced lattice consists on a periodic tiling of the plane by rhombi and is the dual of the kagome lattice. The Potts model on the two-dimensional diced lattice has interesting critical properties. For instance, it has been shown recently \cite{kotecky:030601} that, even though it admits a height mapping representation for $q=3$, it contradicts theoretical expectations having $q_c(diced)>3$. Furthermore, it is the first known case of a two-dimensional bipartite lattice with $q_c>3$.

\begin{figure}[t]
\begin{center}
\includegraphics[width=0.5\textwidth]{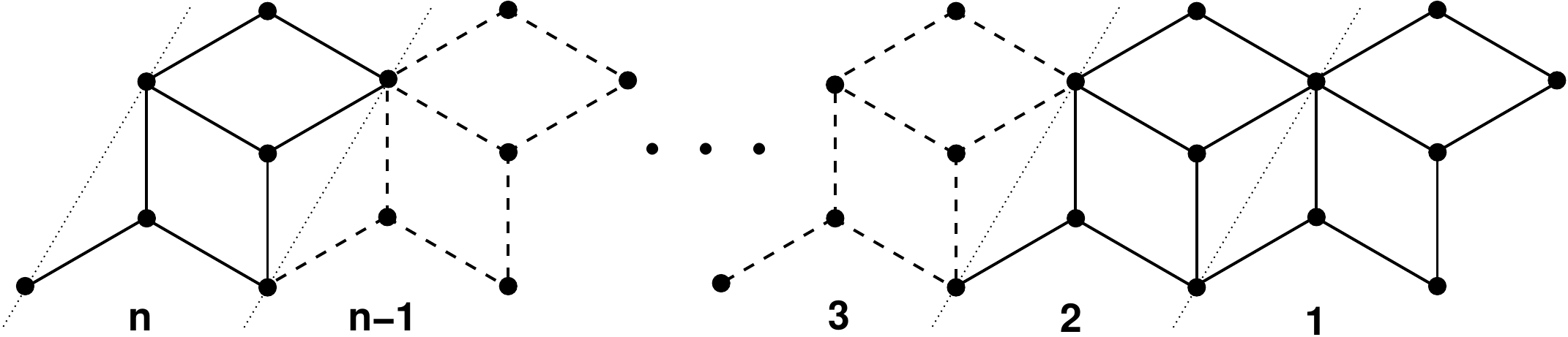}~~~~~~~~~~~~
\end{center}
\caption{A strip of the diced lattice with $L_y=2$. Each unit cell is composed of five vertices and eight edges. A dashed line marks the division between unit cells.}
\label{diced_strip}
\end{figure}

We calculate here the exact partition function for the $L_y=2$ diced lattice strip shown in Fig. \ref{diced_strip}.
Proceeding as before, applying the deletion-contraction theorem to all the vertices in a block, the elements of the resulting two by two transfer matrix $\mathbf{A}^{\text{d}}(q,v)$ are,
\begin{eqnarray}
a_{11}^{\text{d}}
&=& q^5 + 8 q^4 v + 27 q^3 v^2 + 50 q^2 v^3 + 52 q v^4 + 2 q^2 v^4 + 26 v^5 + 7 q v^5 + 9 v^6 + v^7,
\label{d2} \\
a_{12}^{\text{d}}
&=& q^4 v^2 + 6 q^3 v^3 + 16 q^2 v^4 + 22 q v^5 + q^2 v^5 + 15 v^6 + 4 q v^6 + 7 v^7 + v^8,
\label{d3} \\
a_{21}^{\text{d}}
&=& q^4 + 8 q^3 v + 26 q^2 v^2 + q^3 v^2 + 44 q v^3 + 6 q^2 v^3 + 33 v^4 + 19 q v^4 + 26 v^5 + 2 q v^5 + 8 v^6 + v^7,
\label{d4} \\
a_{22}^{\text{d}}
&=& q^3 v^2 + 6 q^2 v^3 + 17 q v^4 + q^2 v^4 + 20 v^5 + 8 q v^5 + 19 v^6 + q v^6 + 7 v^7 + v^8.
\label{d5}
\end{eqnarray}
To analyze the results we plot the internal energy per site and specific heat as a function of the temperature variable $v$, for different values of $q$.

In general, internal energy and specific heat are quantities that can be obtained straightforwardly from the free energy by taking the appropriate derivatives as follows:
\begin{equation}
E = -\frac{\partial f}{\partial \beta} = -J(v+1)\frac{\partial f}{\partial v}
\end{equation}
and
\begin{equation}\label{spec_heat}
\frac{C}{k_B} = \frac{1}{k_B}\frac{\partial E}{\partial T} = K^2 (v+1)\left[
\frac{\partial f}{\partial v} + (v+1)\frac{\partial^2 f}{\partial v^2}\right].
\end{equation}
In the limit of infinite length the free energy per site becomes
\begin{equation}
f = \frac{1}{n_{cell}} \ln \lambda_+,
\end{equation}
where $n_{cell}=n(H)-L_y$ is the number of sites per unit cell and $\lambda_+$ is the dominant eigenvalue with nontrivial coefficient associated. For simplicity, instead of the internal energy, we will plot the dimensionless quantity $-E/J$, which has the virtue of having the same sign for both the ferromagnetic ($0 \leq v\leq \infty$) and antiferromagnetic ($-1 \leq v \leq 0$) physical regions.

\begin{figure}[h]
\begin{center}
\includegraphics[width=0.35\textwidth]{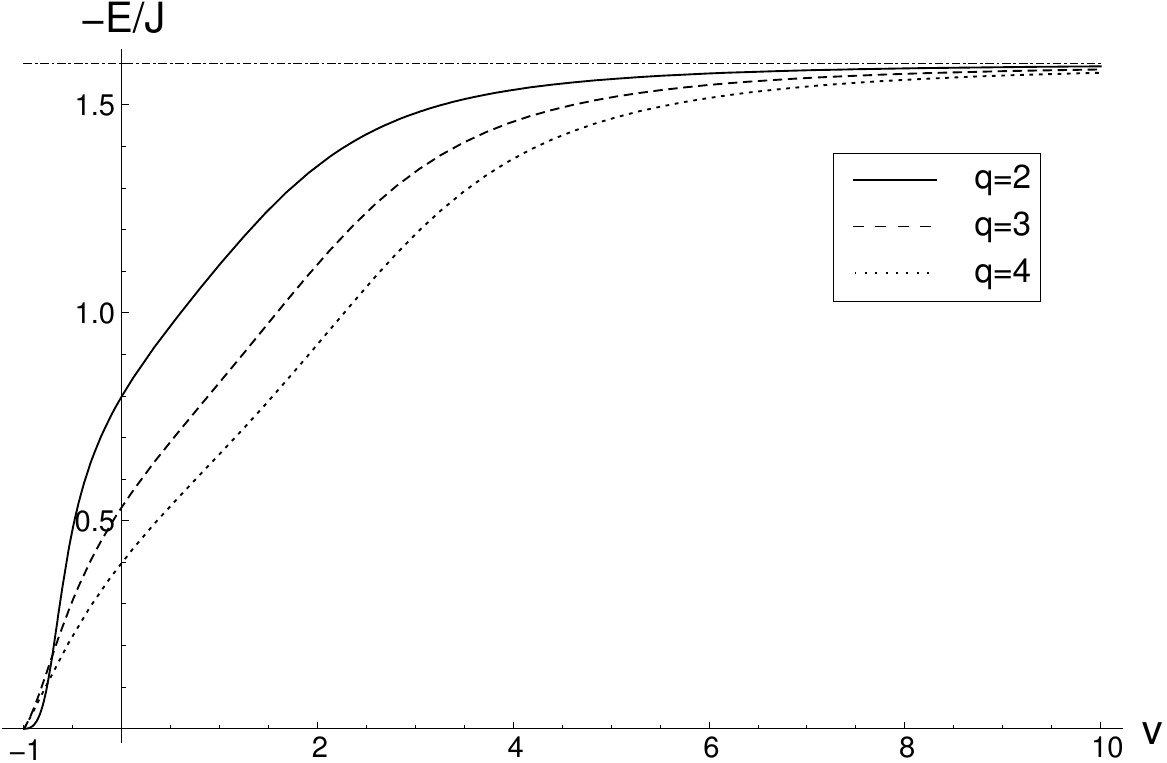}~~~~~~~~~
\includegraphics[width=0.35\textwidth]{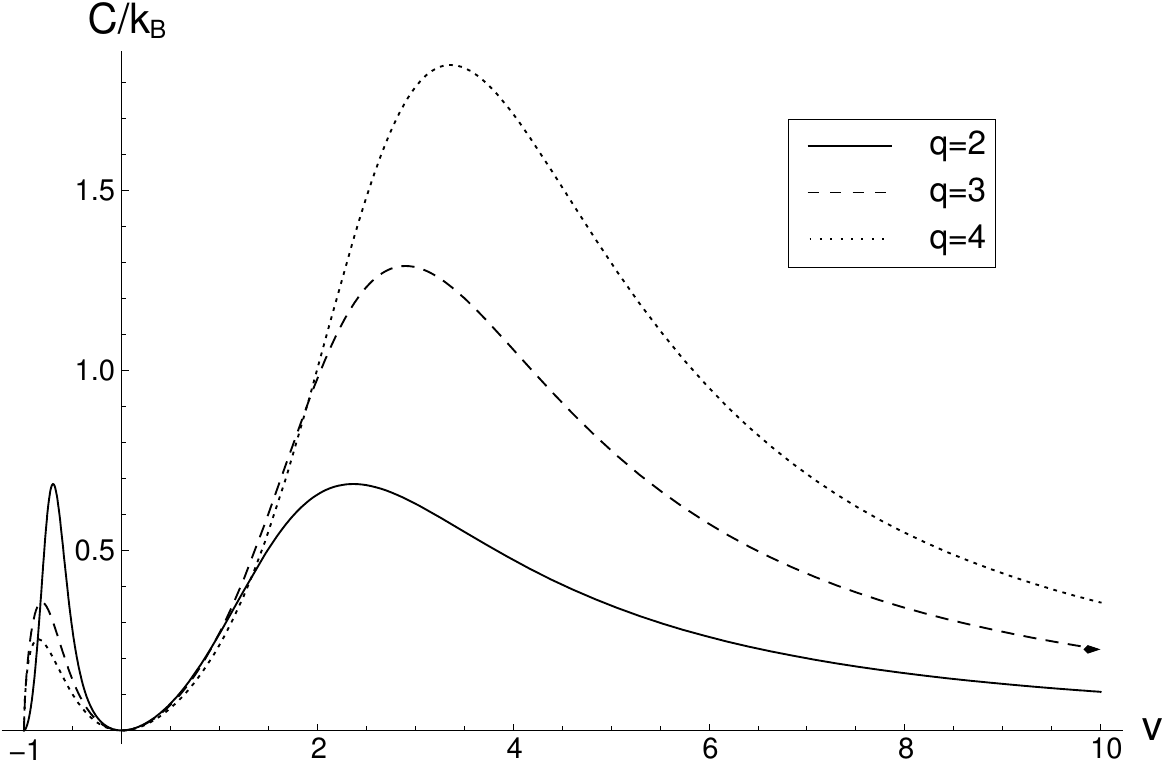}\\[0pt]
(a) \hspace{9cm} (b)
\end{center}
\caption{Results for the diced lattice strip. (a) Reduced internal energy as a function of the temperature parameter $v=e^K-1$ for different values of $q$. (b) Dimensionless specific heat as a function of $v$.}
\label{diced_plots}
\end{figure}

We can now compute the internal energy and specific heat (see Fig. (\ref{diced_plots})).

In this case there is no frustration and even for $q=2$ we observe that the energy reaches $E=0$ at zero temperature. On the ferromagnetic side, the spins become aligned at zero temperature and the minimal value  of energy per site $E=-8J/5$, is reached.

For $q=2$ the model has been solved exactly in two dimensions and it is known to present an antiferromagnetic transition at $v\approx-0.3401$ \cite{Syozi_72}.  Thus, we expect that the observed peak in the specific heat will evolve into the corresponding divergence for infinite width. On the other hand, since for $q=3$ and $4$ the lattice is known to have no antiferromagnetic phase transition, we expect that for such values of $q$ the specific heat peaks at $v<0$ will disappear or move below $v=-1$ as the strip widens.

\subsection{``Shortest-path" lattice strip}\label{sh_path}

\begin{figure}[t]
\begin{center}
\includegraphics[width=0.4\textwidth]{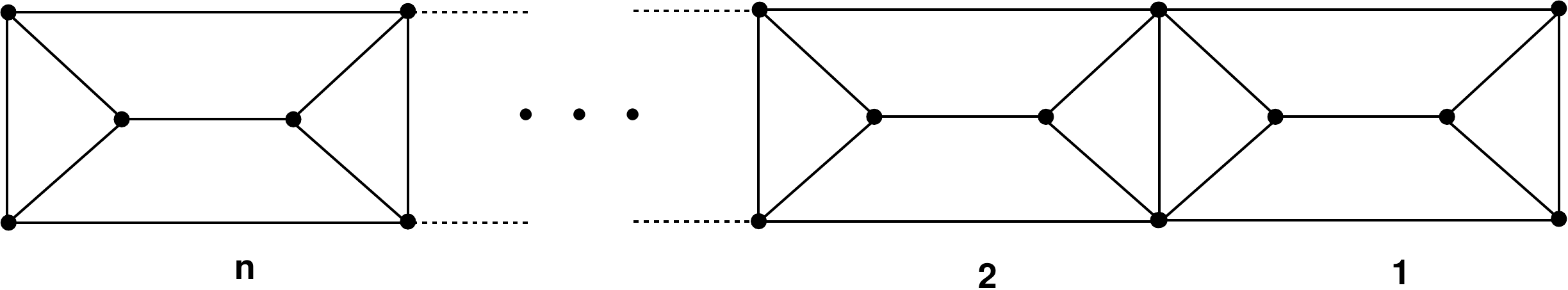}~~~~~~~~~~~~
\end{center}
\caption{A strip of the ``shortest path" lattice with $L_y=2$. Each unit cell is composed of four vertices and eight edges.}
\label{shpath_strip}
\end{figure}

\begin{figure}[h]
\begin{center}
\includegraphics[width=0.4\textwidth]{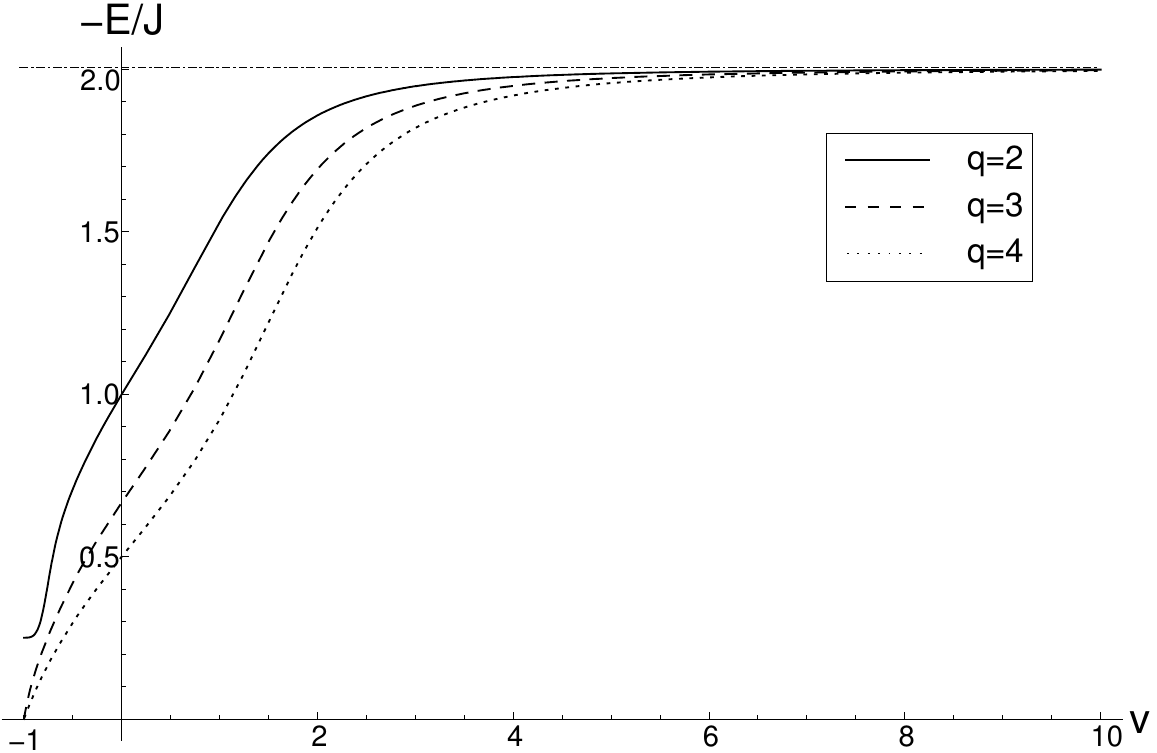}~~~~~~~~~
\includegraphics[width=0.4\textwidth]{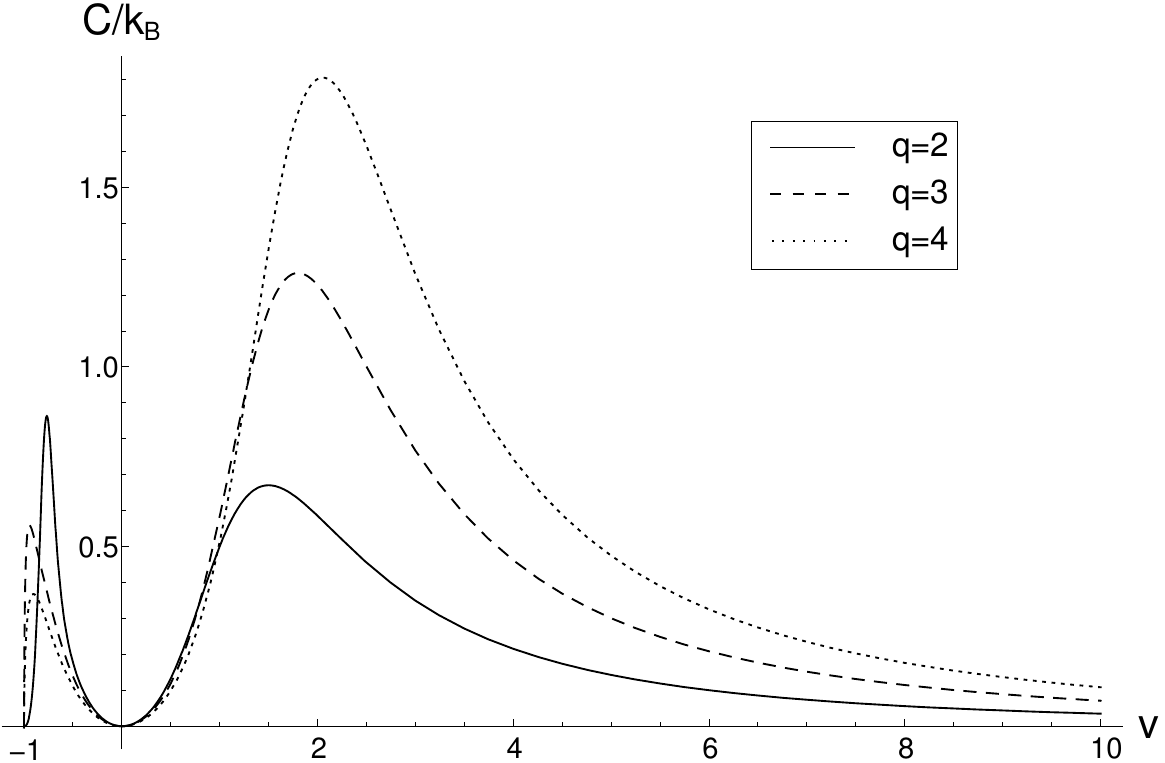}\\[0pt]
(a) \hspace{9cm} (b)
\end{center}
\caption{Results for the ``shortest path" lattice strip. (a) Reduced internal energy as a function of the temperature parameter $v=e^K-1$ for different values of $q$. (b) Dimensionless specific heat as a function of $v$.}
\label{shpath_plots}
\end{figure}

Here we will consider the case of a lattice strip which we will call ``shortest path", since its geometry mimics the topology that naturally appears when one solves the problem of the shortest path connecting the vertices of a square (see Fig. \ref{shpath_strip}).
The recursive equation method yields the following result for the elements of the transfer matrix $\mathbf{A}^{\text{sh}}(q,v)$,
\begin{eqnarray}
a_{11}^{\text{sh}}
&=& q^4 + 8 q^3 v + 27 q^2 v^2 + 48 q v^3 + q^2 v^3 + 40 v^4 + 7 q v^4 + 16 v^5 + 2 v^6,
\label{sh2} \\
a_{12}^{\text{sh}}
&=& q^3 v^2 + 7 q^2 v^3 + 23 q v^4 + 35 v^5 + 5 q v^5 + 26 v^6 + 8 v^7 + v^8,
\label{sh3} \\
a_{21}^{\text{sh}}
&=& q^3 + 7 q^2 v + q^3 v + 18 q v^2 + 8 q^2 v^2 + 20 v^3 + 23 q v^3 + q^2 v^3 + 32 v^4 + 5 q v^4 + 14 v^5 + 2 v^6,
\label{sh4} \\
a_{22}^{\text{sh}}
&=& 2 q^2 v^2 + 10 q v^3 + 2 q^2 v^3 + 20 v^4 + 13 q v^4 + 39 v^5 + 3 q v^5 + 26 v^6 + 8 v^7 + v^8.
\label{sh5}
\end{eqnarray}
Internal energy and specific heat are plotted in Fig. \ref{shpath_plots} for different values of $q$.

Frustration for $q=2$ is manifested in the finite energy at zero temperature in the antiferromagnetic side. As for the case of the kagome strip, it is easy to find the minimum energy configuration for this system and find that its energy becomes $E=-J/4$, as can be verified in Fig. \ref{shpath_plots}(a).

Although to our knowledge this lattice has not been studied before, due to universality in the ferromagnetic side it is safe to state that the observed specific heat maxima at $v>0$ are related to the divergences at the transitions expected in two dimensions. On the other hand, the Potts antiferromagnet is highly dependent on its lattice structure and extracting information about its critical properties would require further study. \\

\subsection{Some remarks on Fisher zeros}\label{App_A}

It is sometimes useful to find the zeros of the partition function in the complex $v$ plane. These are the so called Fisher zeros which have been used to develop very powerful tools in statistical mechanics \cite{PhysRev.87.404,PhysRev.87.410,PTP.37.1070,PTP.38.72,PTP.38.289,PTP.39.349}. We have analyzed the Fisher zeros of the square, diced, ``shortest-path" and kagome strip lattices for three ranges of values of $q$: small-$q$ limit, small integers (lets say 2,3,4) and the large-$q$ limit.

\begin{figure}[h]
\begin{center}
\includegraphics[width=0.27\textwidth]{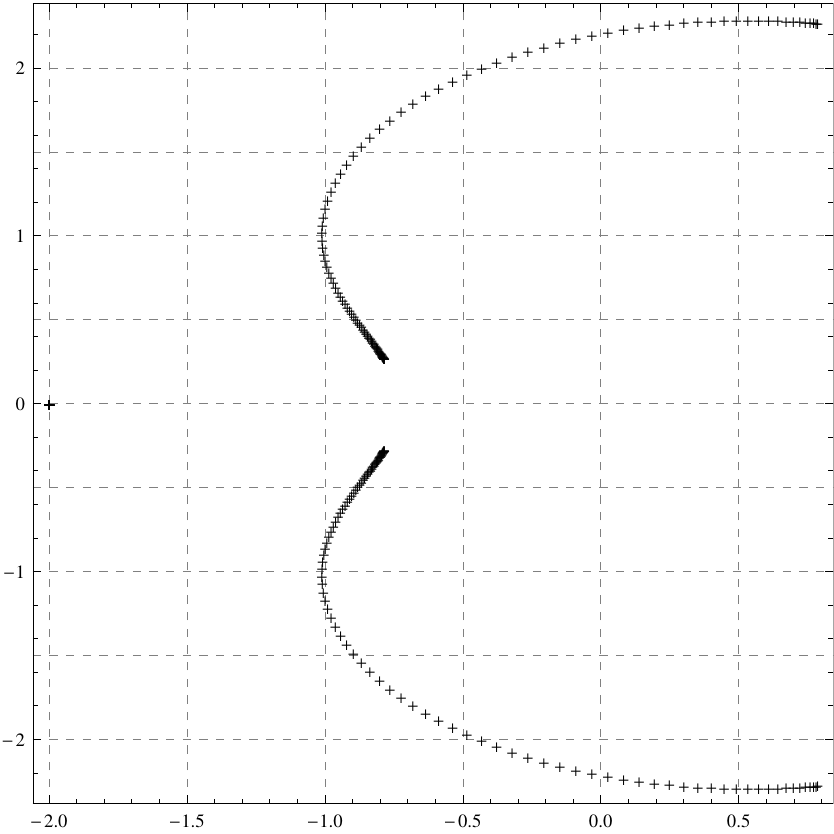}~~~~~~~~~~~~
\includegraphics[width=0.27\textwidth]{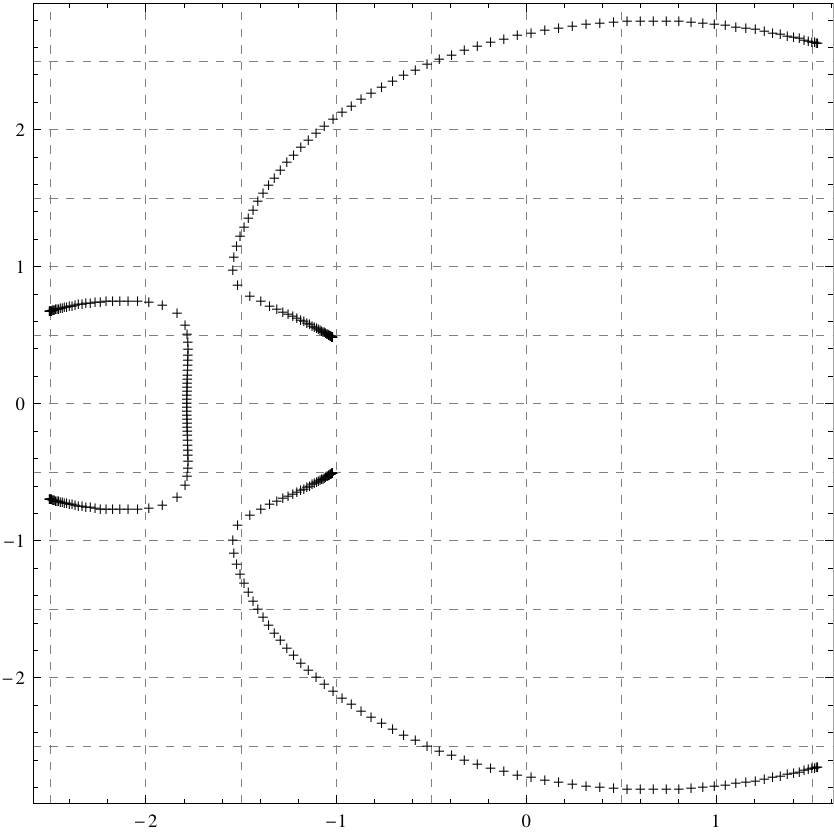}~~~~~~~~~~~~
\includegraphics[width=0.27\textwidth]{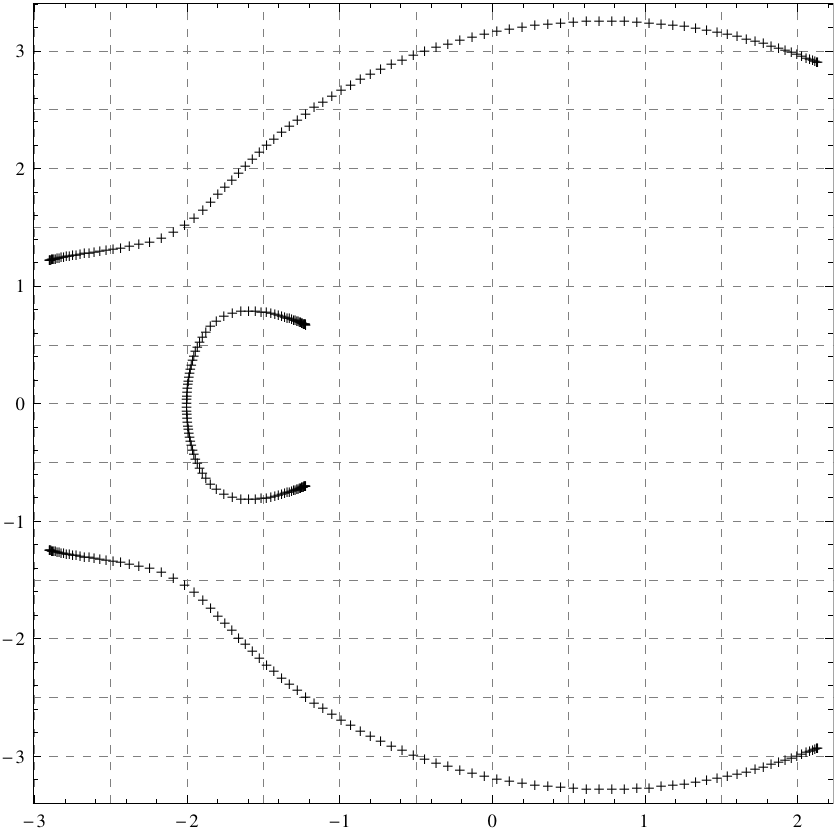}
\end{center}
\hspace{2.4cm}(a1)\hspace{5.6cm}(b1)\hspace{5.8cm}(c1)
\begin{center}
\includegraphics[width=0.27\textwidth]{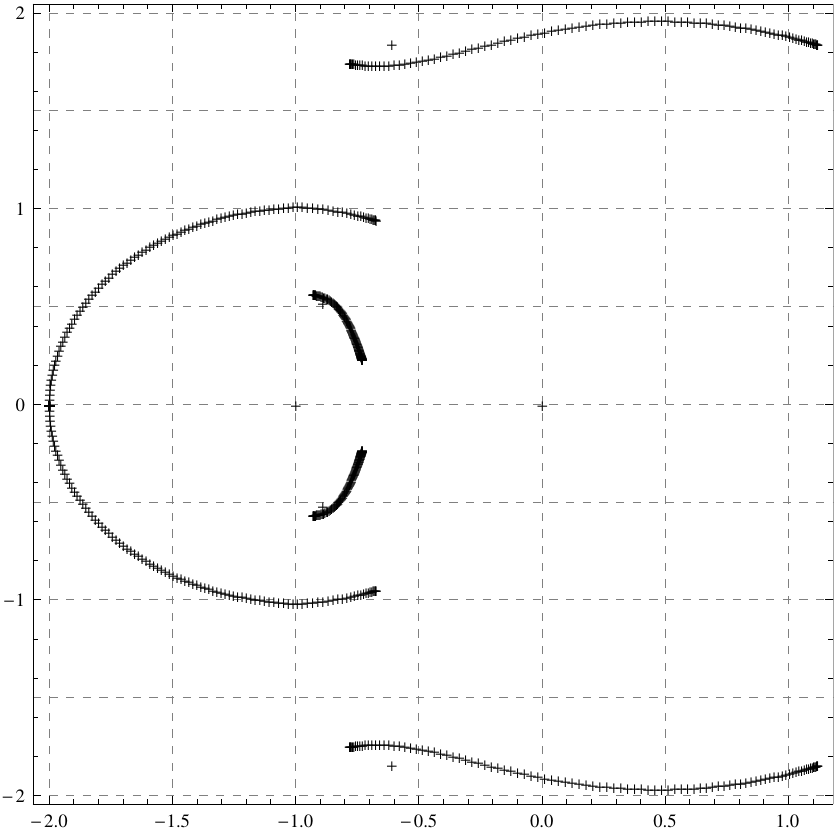}~~~~~~~~~~~~
\includegraphics[width=0.27\textwidth]{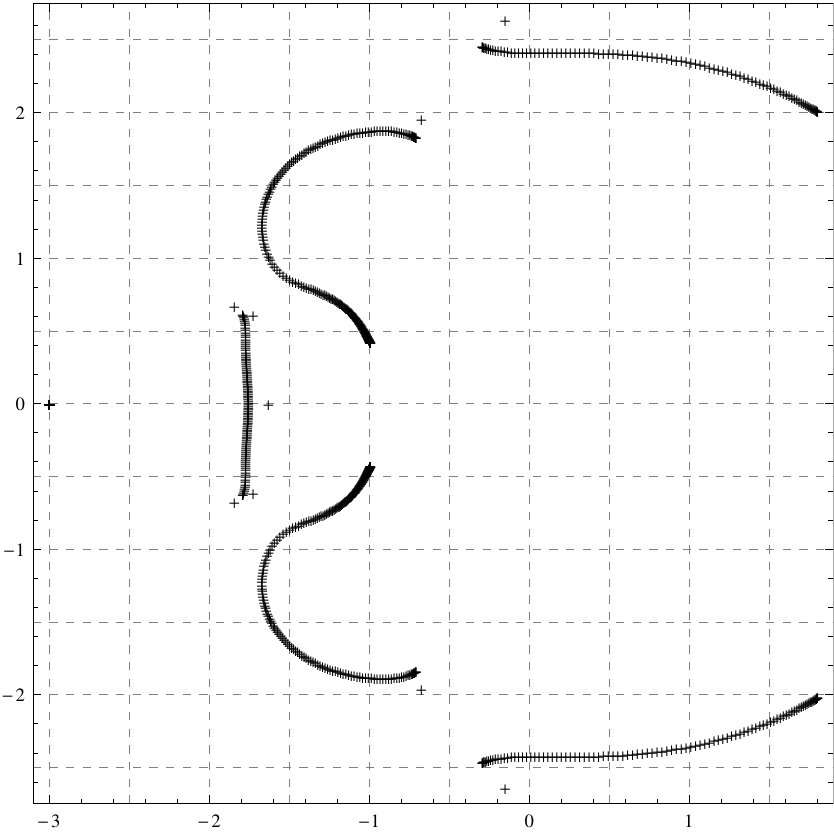}~~~~~~~~~~~~
\includegraphics[width=0.27\textwidth]{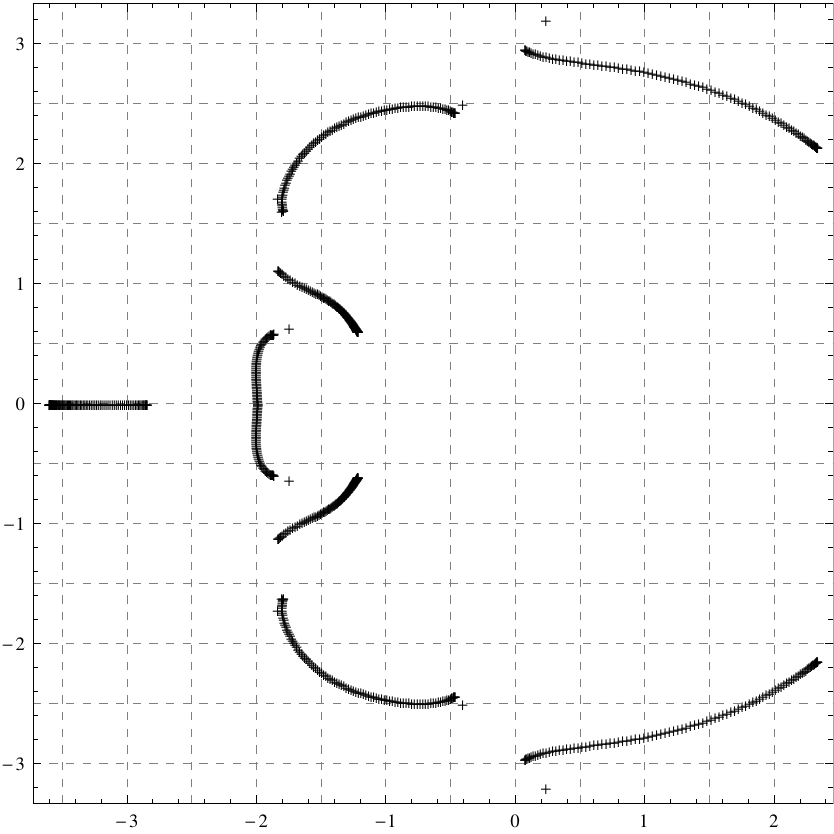}
\end{center}
\hspace{2.4cm}(a2)\hspace{5.6cm}(b2)\hspace{5.8cm}(c2)
\begin{center}
\includegraphics[width=0.27\textwidth]{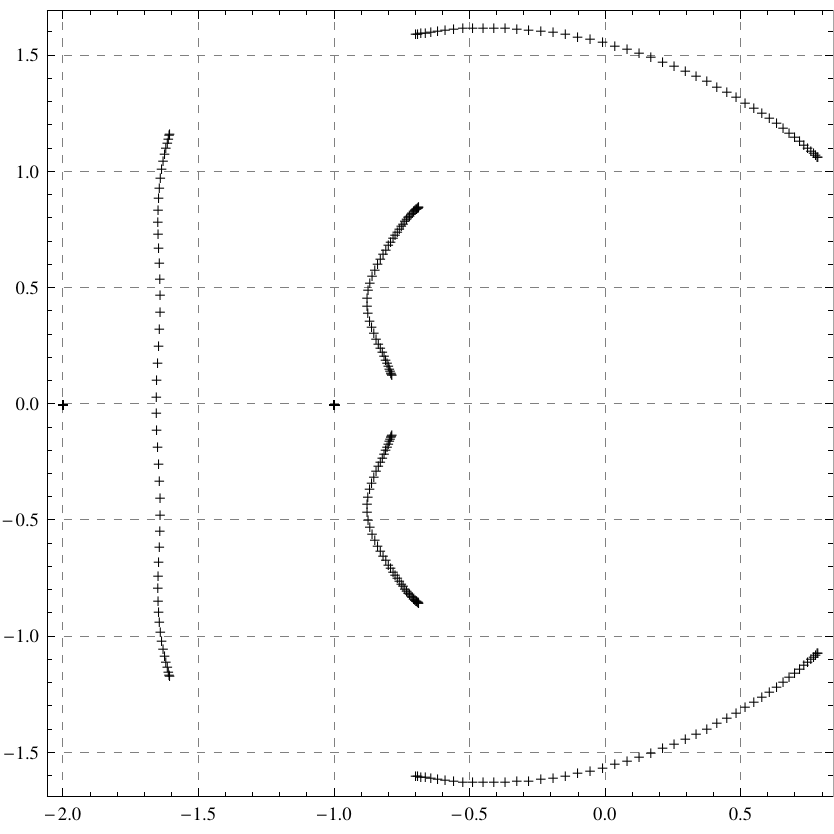}~~~~~~~~~~~~
\includegraphics[width=0.27\textwidth]{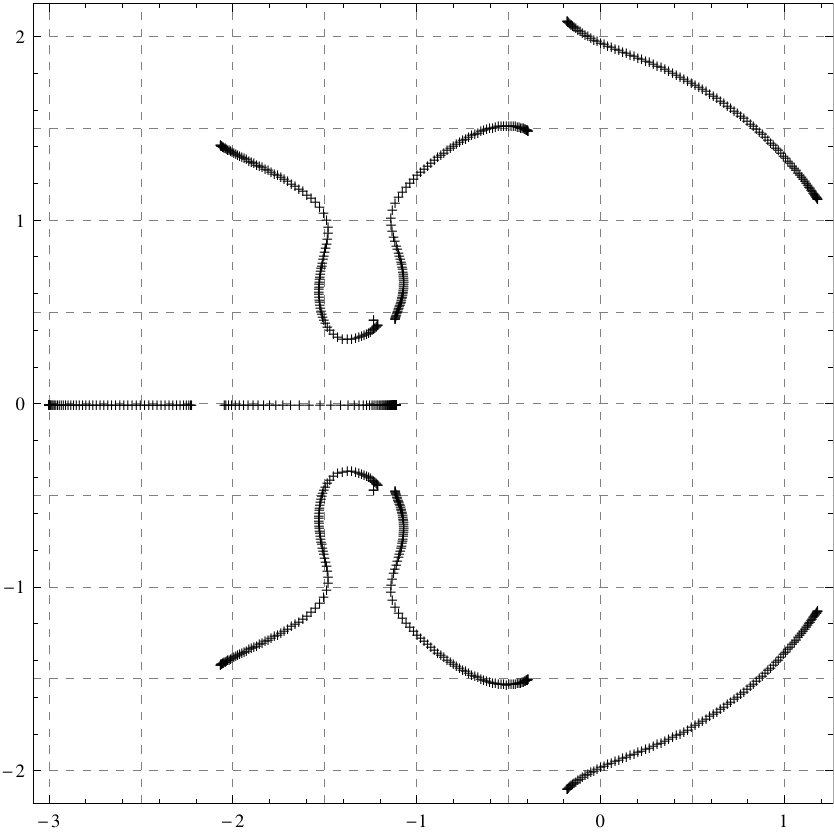}~~~~~~~~~~~~
\includegraphics[width=0.27\textwidth]{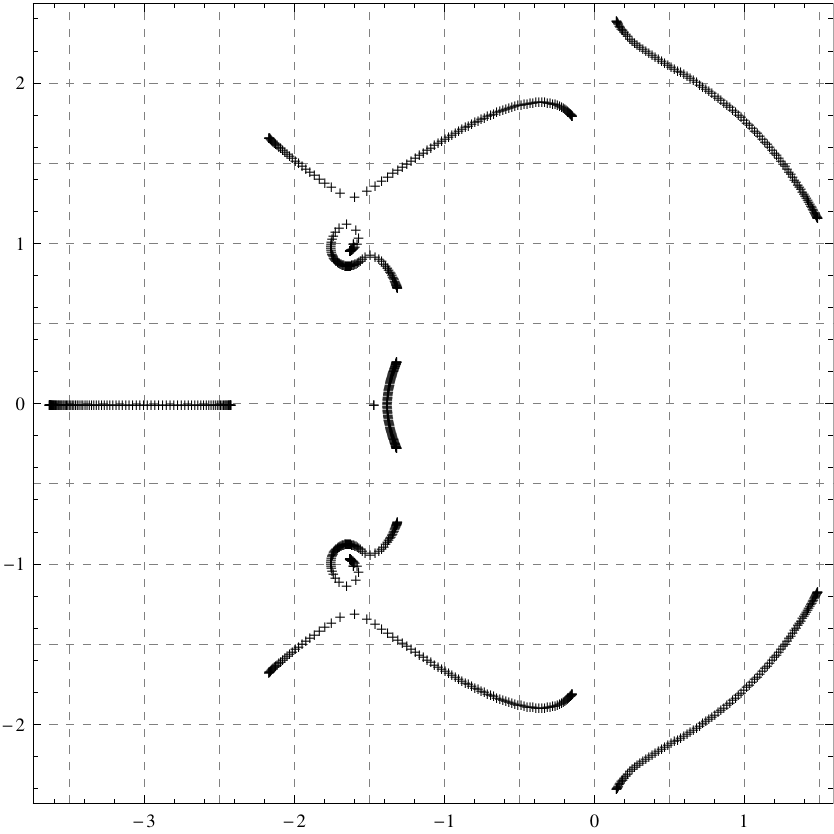}
\end{center}
\hspace{2.4cm}(a3)\hspace{5.6cm}(b3)\hspace{5.8cm}(c3)
\begin{center}
\includegraphics[width=0.28\textwidth]{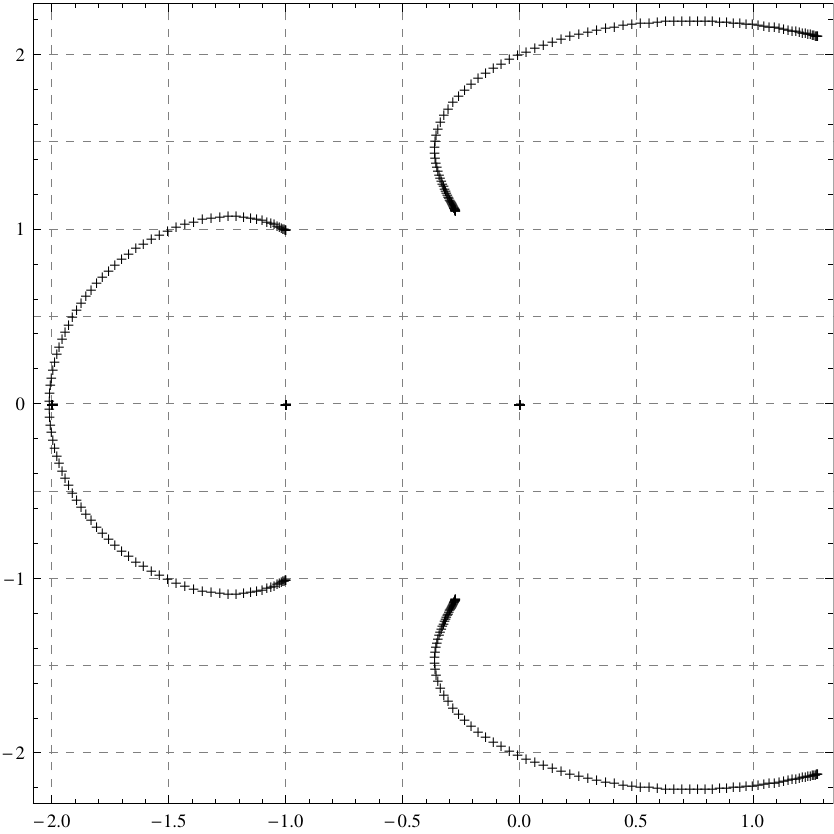}~~~~~~~~~~~~
\includegraphics[width=0.28\textwidth]{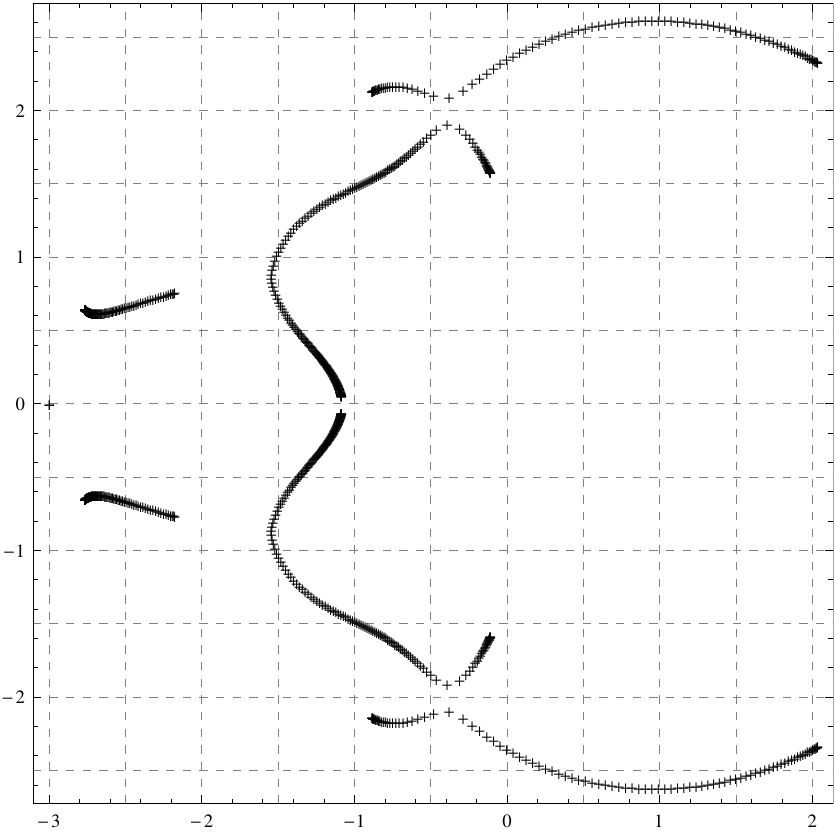}~~~~~~~~~~~~
\includegraphics[width=0.28\textwidth]{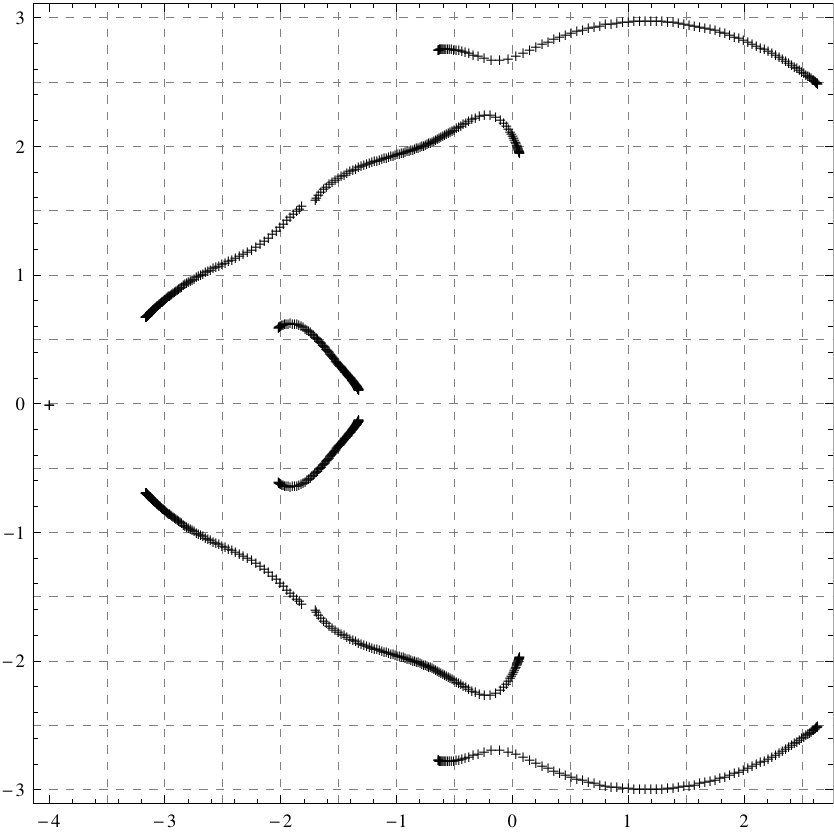}
\end{center}
\hspace{2.4cm}(a4)\hspace{5.6cm}(b4)\hspace{5.8cm}(c4)
\caption{ The Fisher zeros in the complex $v$ plane for strips of length $n=100$. In columns there are the values: (a) $q=2$, (b) $q=3$, and (c) $q=5$; and in rows lattices (1) square, (2) diced, (3) ``shortest-path" and (4) kagome. Note: only (a3) is for $n=50$.} \label{Fisher_st}
\end{figure}

When $q$ is in the small integer range the structure of the Fisher zeros for the two lattices is very sensible to the structure of the lattice as illustrated in Fig. \ref{Fisher_st}. On the other hand, for small-$q$ and large-$q$ limits it can be easily read from the diagrams that with a very good approximation the zeros are given by the zeros of the functions
\begin{equation}\label{small&large-q}
F_{q\rightarrow0}=q(v+1)^{n_e}\,, \qquad F_{q\rightarrow\infty}=q^{n_v}+q v^{n_e}\,,
\end{equation}
where $n_e$ is the total number of edges of the strip and $n_v$ is the total number of vertices. The actual plots for the Fisher zeros in the small-$q$ and large-$q$ limit are not displayed here because they are just a dot and a circle given by eqs. (\ref{small&large-q}). Within the current conventions the formulas for $n_e$ and $n_v$ for strips of width $L_y=2$ are given in the table \ref{table1}.

\begin{table}[h]
\begin{center}
\begin{tabular}{|l|c|c|}
\hline
unit cell & $n_e$ & $n_v$\\
\hline
square            & $3n-2$ & $2n$\\
diced             & $8n-7$ & $5n-3$\\
''shortest-path'' & $8n-7$ & $4n-2$\\
kagome            & $8(n-1)$ & $5n-3$\\
\hline
\end{tabular}
\caption{Total number of edges and vertices for the lattices of width $2\times n$ presented here.}\label{table1}
\end{center}
\end{table}

\section{Kagome lattice strip $L_y=2$ to $5$}\label{kagome}

The kagome lattice is the medial graph of both the triangular and hexagonal lattices. The Potts model on this network can exhibit nonzero entropy at zero temperature with or without frustration depending of the value of $q$ and has been widely studied. Nevertheless, in contrast with simpler lattices like the square, triangular or honeycomb \cite{Baxter73,Baxter78,Hintermann78}, many fundamental questions such as the exact determination of the critical frontier remain unresolved in the kagome network \cite{Wu79,HU94,Chen99,PhysRevE.67.017103,Feldmann:1998zz}. In this context, exact results for kagome strips of finite width may provide some insight into the properties of the model in two dimensions.

\begin{figure}[t]
\begin{center}
\includegraphics[width=0.4\textwidth]{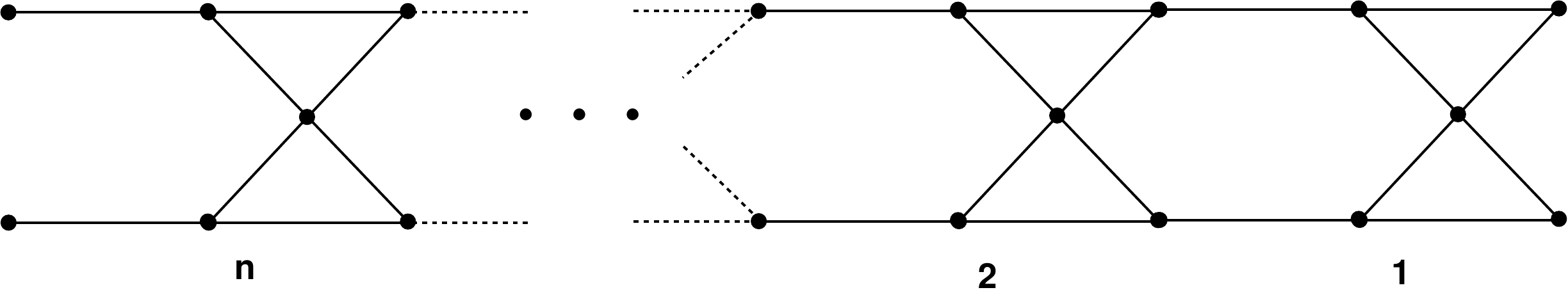}~~~~~~~~~~~~
\end{center}
\caption{A strip of the kagome lattice with $L_y=2$. }
\label{kag_strip}
\end{figure}

\begin{figure}[h]
\begin{center}
\includegraphics[width=0.4\textwidth]{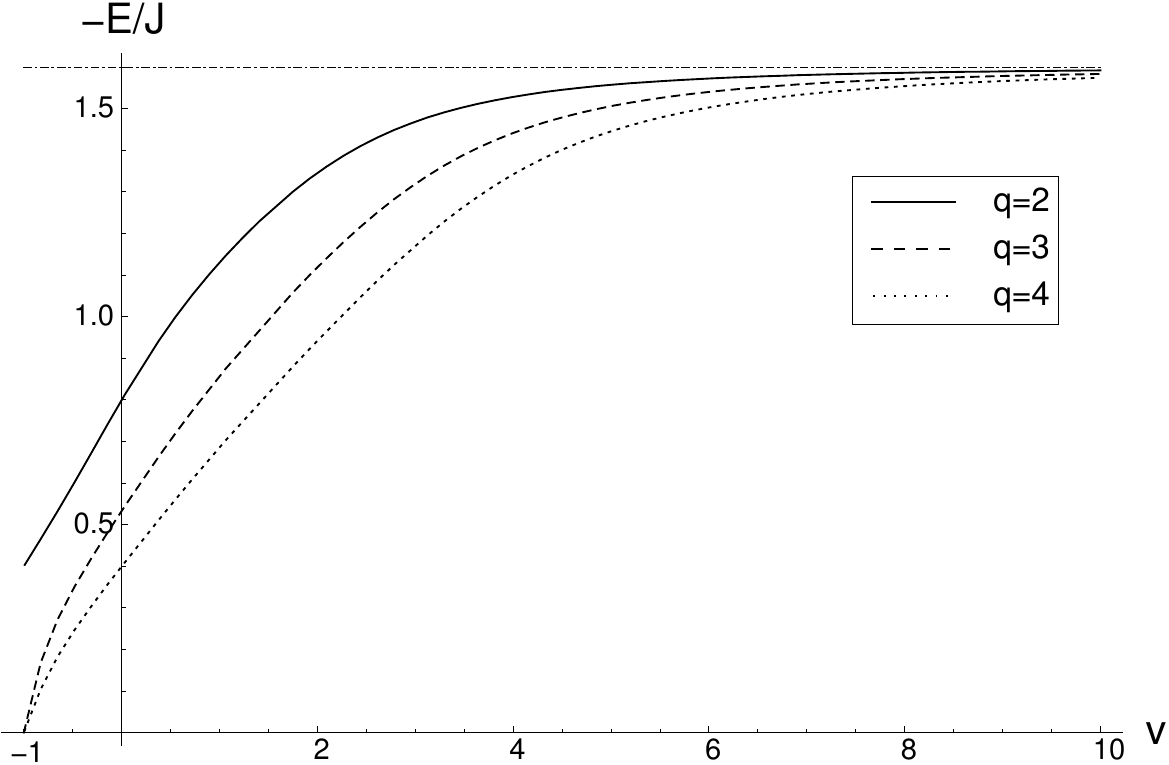}~~~~~~~~~
\includegraphics[width=0.4\textwidth]{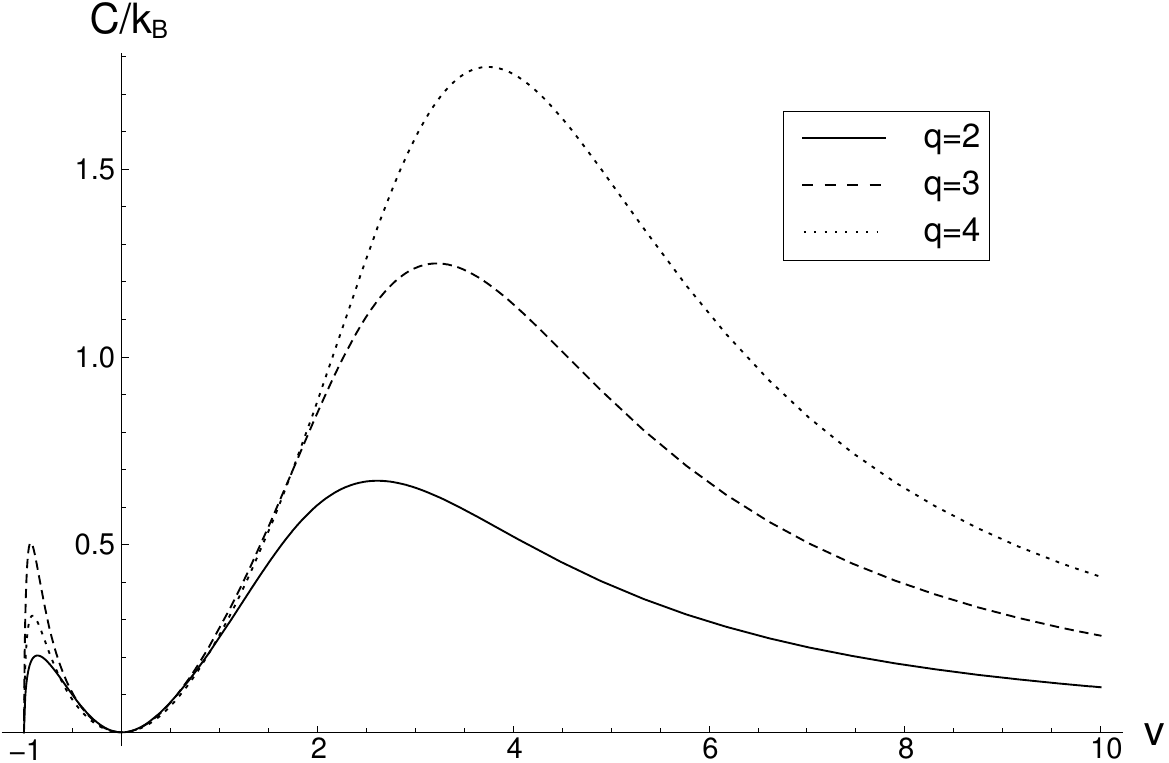}\\[0pt]
(a) \hspace{9cm} (b)
\end{center}
\caption{Results for the kagome lattice strip. (a) Reduced internal energy as a function of the temperature parameter $v=e^K-1$ for different values of $q$. Lattice frustration gives rise to a finite zero-temperature internal energy for $q=2$. (b) Dimensionless specific heat as a function of $v$.}
\label{kag_plots}
\end{figure}

First, we proceed to use the recursive equation method to calculate the exact partition function for the $L_y=2$ kagome strip shown in Fig. \ref{kag_strip}.
The elements of the transfer matrix are,
\begin{eqnarray}
a_{11}^{\text{k}} &=& q^5 + 8 q^4 v + 28 q^3 v^2 + 54 q^2 v^3 + 2 q^3 v^3 + 59 q
v^4 + 10 q^2 v^4 + 30 v^5 + 20 q v^5 + 16 v^6 + q v^6 + 2 v^7,  \label{k2} \\
a_{12}^{\text{k}} &=& q^2 v^4 + 6 q v^5 + 9 v^6 + 2 q v^6 + 6 v^7 + v^8, \label{k3} \\
a_{21}^{\text{k}} &=& q^4 + 8 q^3 v + 27 q^2 v^2 + q^3 v^2 + 46 q v^3 + 10 q^2
v^3 + 36 v^4 + 30 q v^4 + 2 q^2 v^4 \\
&~& + 34 v^5 + 10 q v^5 + 14 v^6 + q v^6 + 2 v^7,  \label{k4} \\
a_{22}^{\text{k}} &=& 2 q v^4 + 12 v^5 + 13 v^6 + 6 v^7 + v^8.  \label{k5}
\end{eqnarray}
It is now straightforward to obtain the eigenvalues of this two by two matrix. They will not be written here explicitly but, as explained above, they are used to calculate the internal energy and specific heat that are displayed in Fig. \ref{kag_plots}.

Due to frustration, a finite zero-temperature internal energy is observed for $q=2$ in the antiferromagnetic regime ($v=-1$). It is actually not hard to find the exact ground state configuration which consists of only two edges per unit cell with energy $-J$. Since there are five sites per unit cell, the energy per site becomes $-2J/5=-0.4J$ as observed in Fig. \ref{kag_plots}a. For larger values of $q$ there is no frustration and the energy goes to zero as $v\rightarrow -1$. At low temperatures on the ferromagnetic side ($v\rightarrow \infty$) the ground state consists always in an alignment of all the spins. Thus, each one of the eight edges in the unit cell will have energy $-J$, resulting in an energy per site of $-8J/5=-1.6J$, as shown in Fig. \ref{kag_plots}a.

The specific heat presents a peak in the ferromagnetic side, which is related to the phase transition known to be present in the 2D kagome lattice. As the strip widens, the peak will evolve into a characteristic discontinuity of the transition. Furthermore, since the Wu conjecture \cite{Wu79} determines with great accuracy that the ferromagnetic critical points are located at $v_c(q=2)\approx 1.542$, $v_c(q=3)\approx 1.876$ and $v_c(q=4)\approx 2.156$ \footnote{The Wu conjecture is known to be wrong, albeit quite precise.}, we know that the position of the peaks will move to the left for wider strips.

As expected, the Potts antiferromagnet is much more complicated. The $q=2$ case on the kagome lattice has been solved exactly \cite{Syozi_72} and it is known to present no phase transition at any temperature. Moreover, it is expected that the $q=3$ model is critical at zero temperature \cite{PhysRevB.45.7536,Kondev96} and that it remains noncritical for any $q>3$. Thus, the peaks for $v<0$ observed in Fig. \ref{kag_plots}b are not expected to evolve into phase transition divergences as the strip widens for $q=2,4$; whereas for $q=3$ we do expect that to be the case. It is interesting to note that indeed the most pronounced maxima corresponds to $q=3$.

So far we have demonstrated the versatility of the deletion-contraction procedure in handling lattice strips of width $L_y=2$. We will now consider wider strips. As the strips become wider, the size of the transfer matrices grows rapidly and performing the corresponding calculations requires the assistance of a computer software. In particular, the size of the transfer matrix for a kagome strip of {\it odd} width $L_y$ under free boundary conditions is given by Catalan's number,
\begin{equation}
 C_{L_y}=\frac{1}{L_y+1}\left(\begin{array}{c}
                             2L_y \\
                             L_y
                            \end{array}
\right),
\end{equation}
which gives the number of non-crossing partitions of the set $\{1,2,...,L_y\}$.\footnote{For {\it even} widths $L_y$ the size is reduced due to reflection symmetry.} For instance, for a kagome strip of width $L_y=5$ we need to calculate a $42 \times 42$ transfer matrix.

To perform such calculations for strips with $L_y>2$ we have developed a code in {\it Mathematica}. Basically, the code performs the same task that one would do by hand: Repeatedly apply the deletion contraction theorem to one block of the strip. It does so for each one of the 42 possible initial graphs and, by keeping track of the coefficients corresponding to the final graphs, it constructs the desired transfer matrix. The outputs are available online in a {\it Mathematica}-friendly format in the electronic archive {\tt arXiv}. The corresponding filenames are {\tt A\_kago\_m=}$L_y${\tt .txt} and {\tt AID\_kago\_m=}$L_y${\tt .txt} ($L_y=2,3,4,5$) for the matrices and the initial data respectively.

\begin{figure}[t]
\begin{center}
\includegraphics[width=0.4\textwidth]{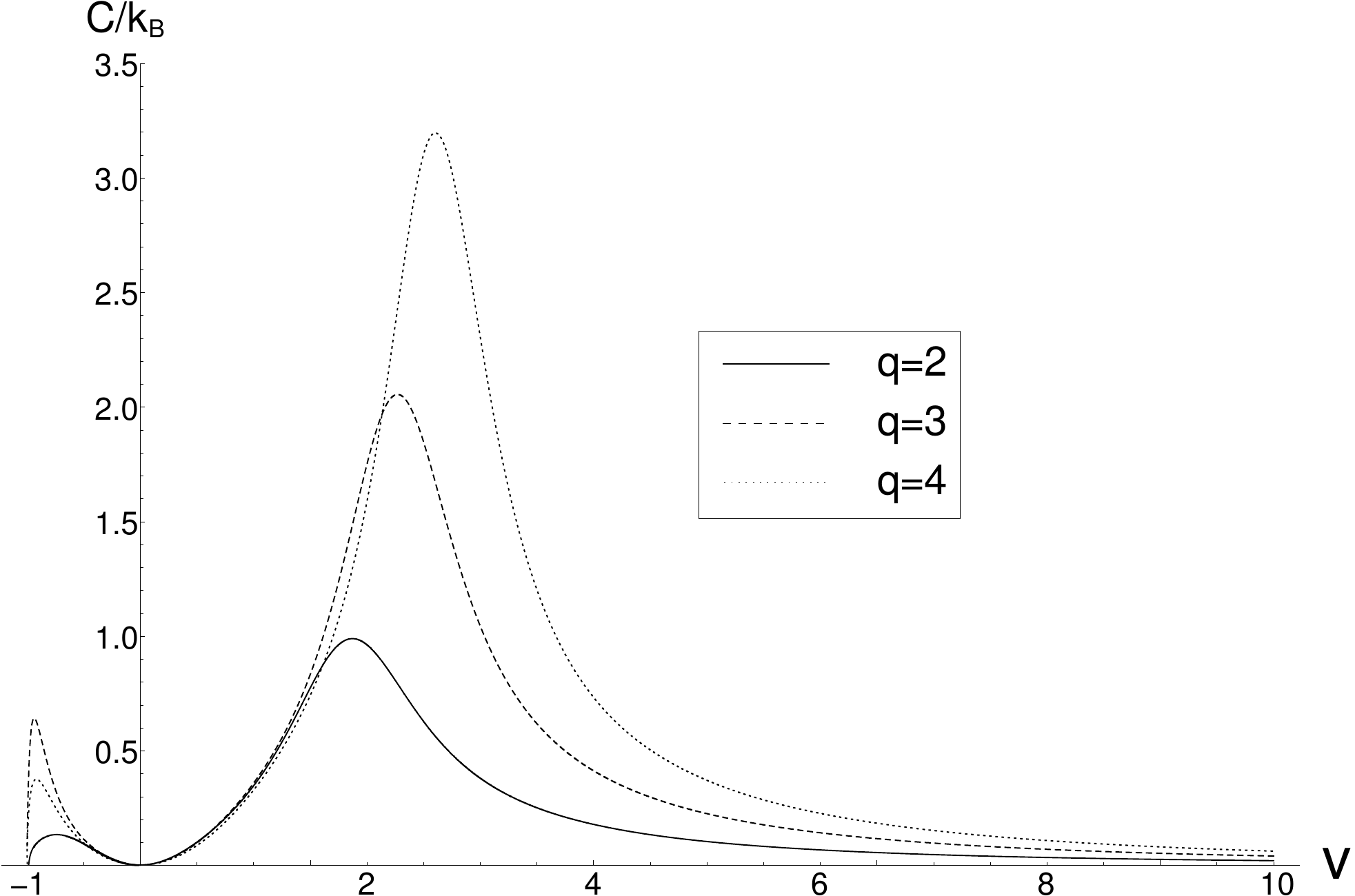}\\[0pt]
\end{center}
\caption{Specific heat for an infinite kagome lattice strip of width $L_y=5$ as a function of $v$.}
\label{spec_heat_Kagome_5}
\end{figure}

As before, the specific heat can be calculated  using (\ref{spec_heat}). The results displayed in Fig. \ref{spec_heat_Kagome_5} show some interesting features. In the ferromagnetic side the maxima become more pronounced as expected for wider strips. Furthermore, all of the peaks come closer to the phase transition points expected for the two dimensional kagome lattice.

\begin{figure}[h]
\begin{center}
\includegraphics[width=0.47\textwidth]{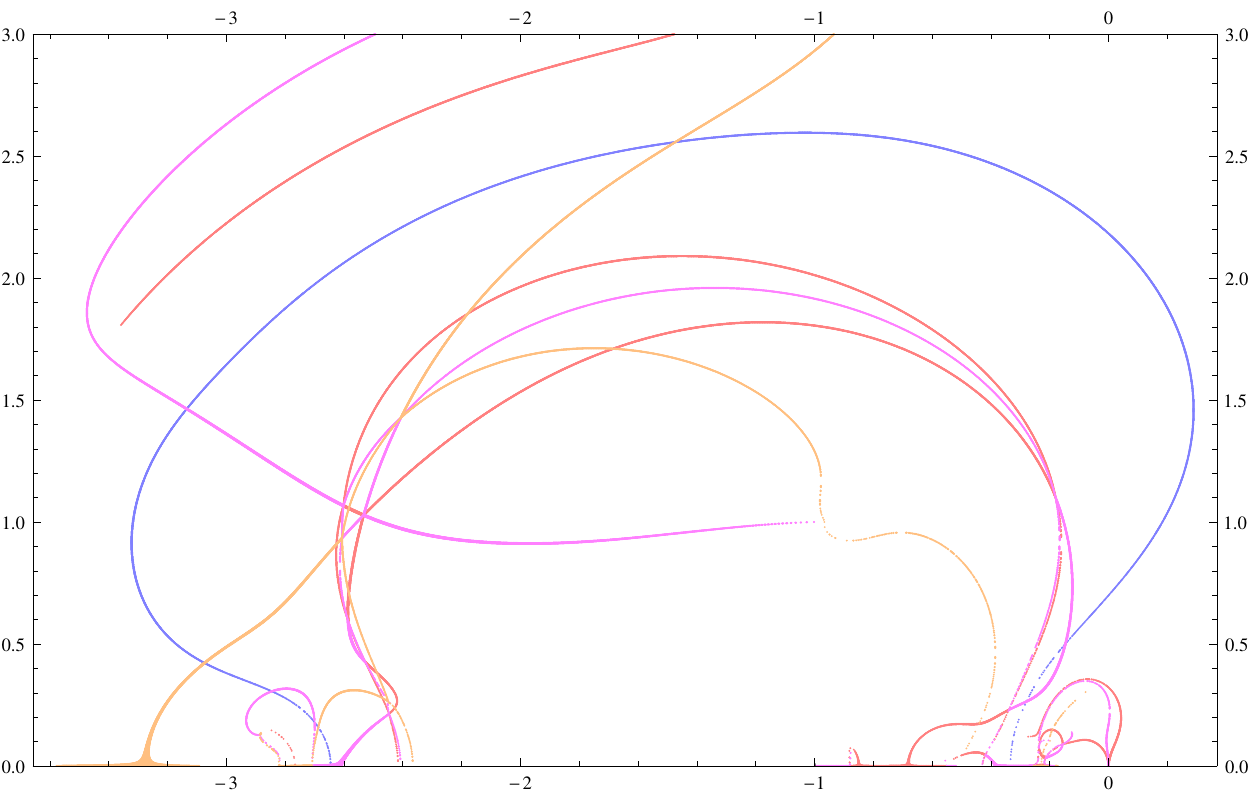}~~~~~
\includegraphics[width=0.47\textwidth]{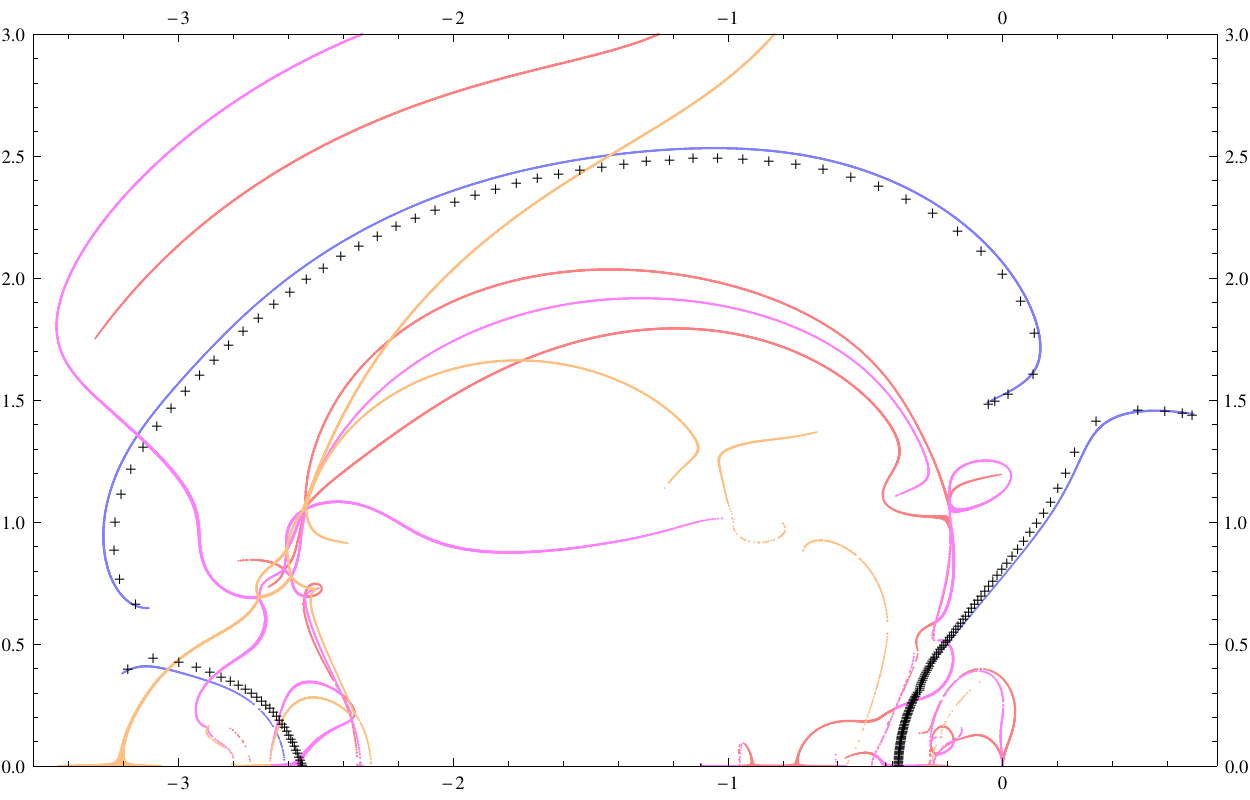}
\end{center}
\hspace{3 cm}(a) $m=3$, $q=1$, \hspace{5.5 cm}(b) $m=3$, $q=1.1$, $n=50$,
\begin{center}
\includegraphics[width=0.47\textwidth]{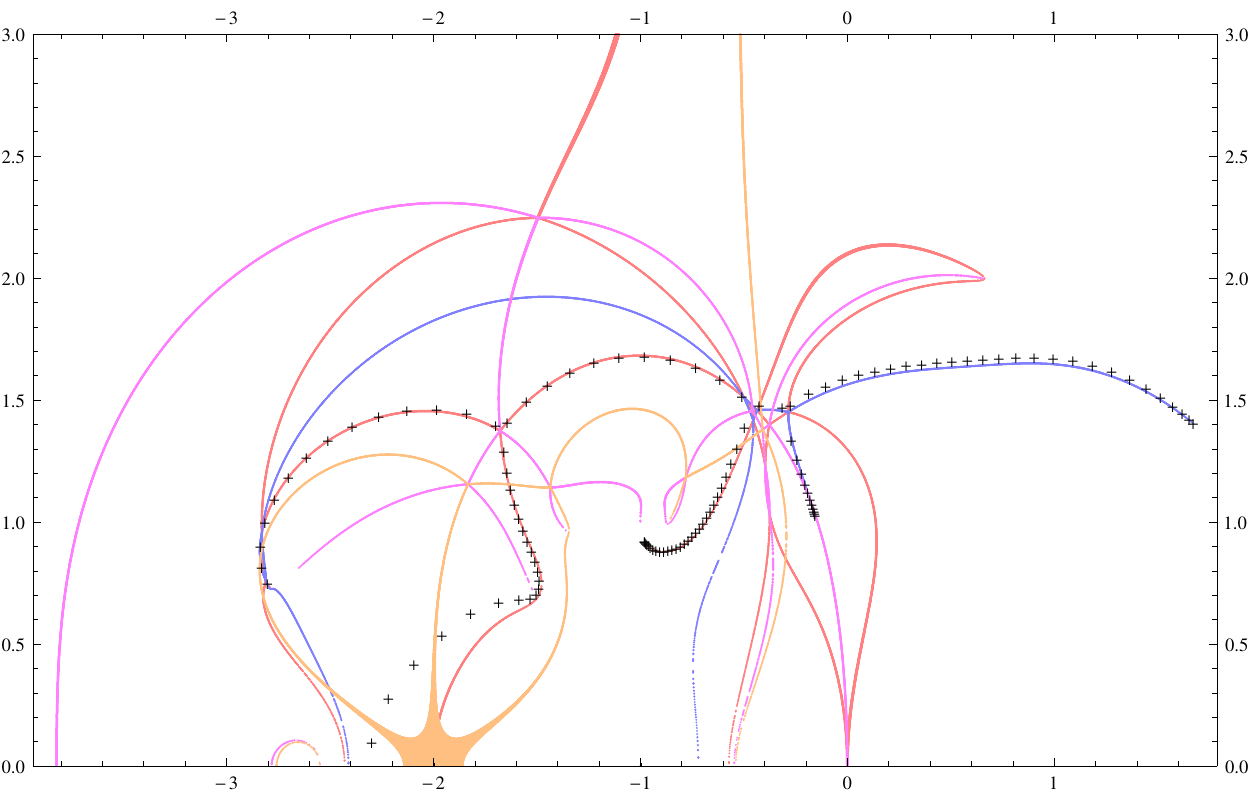}~~~~~
\includegraphics[width=0.47\textwidth]{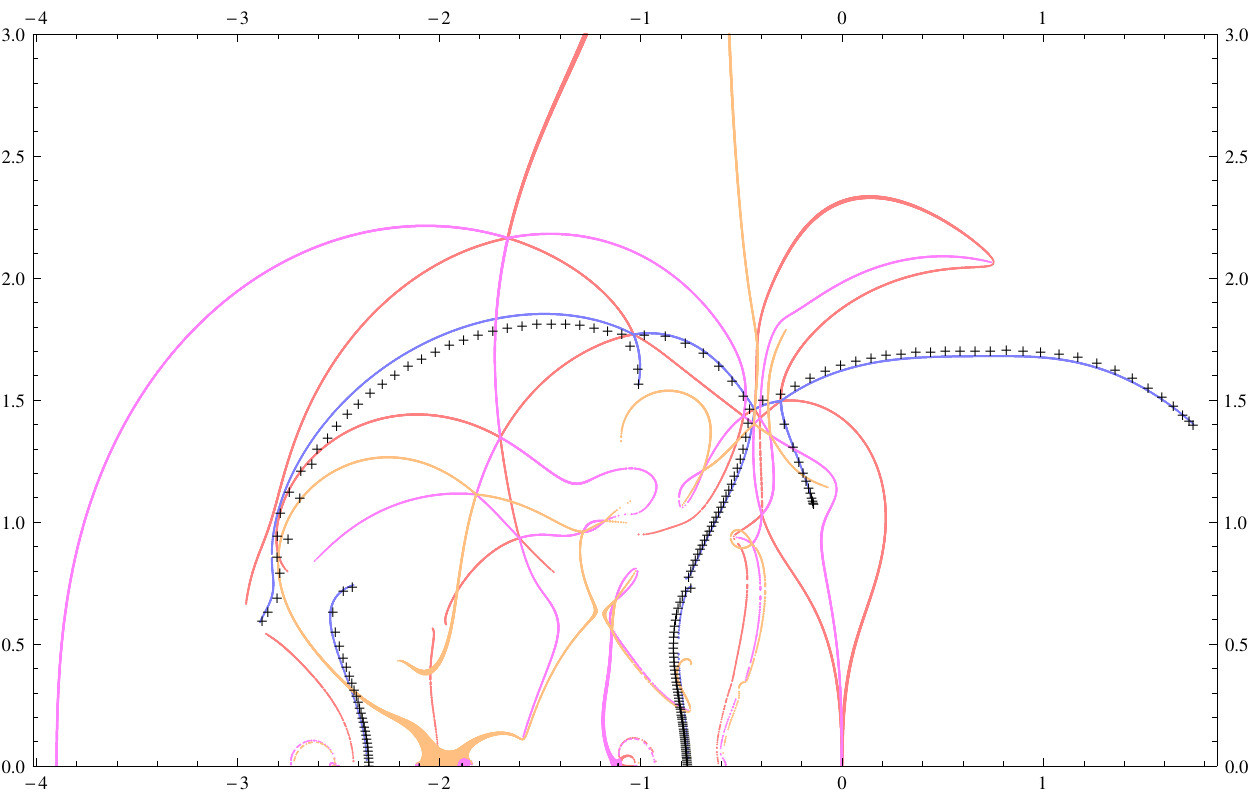}
\end{center}
\hspace{2.5 cm}(c) $m=3$, $q=2$, $n=50$, \hspace{4.5 cm}(d) $m=3$, $q=2.1$, $n=50$,
\begin{center}
\includegraphics[width=0.47\textwidth]{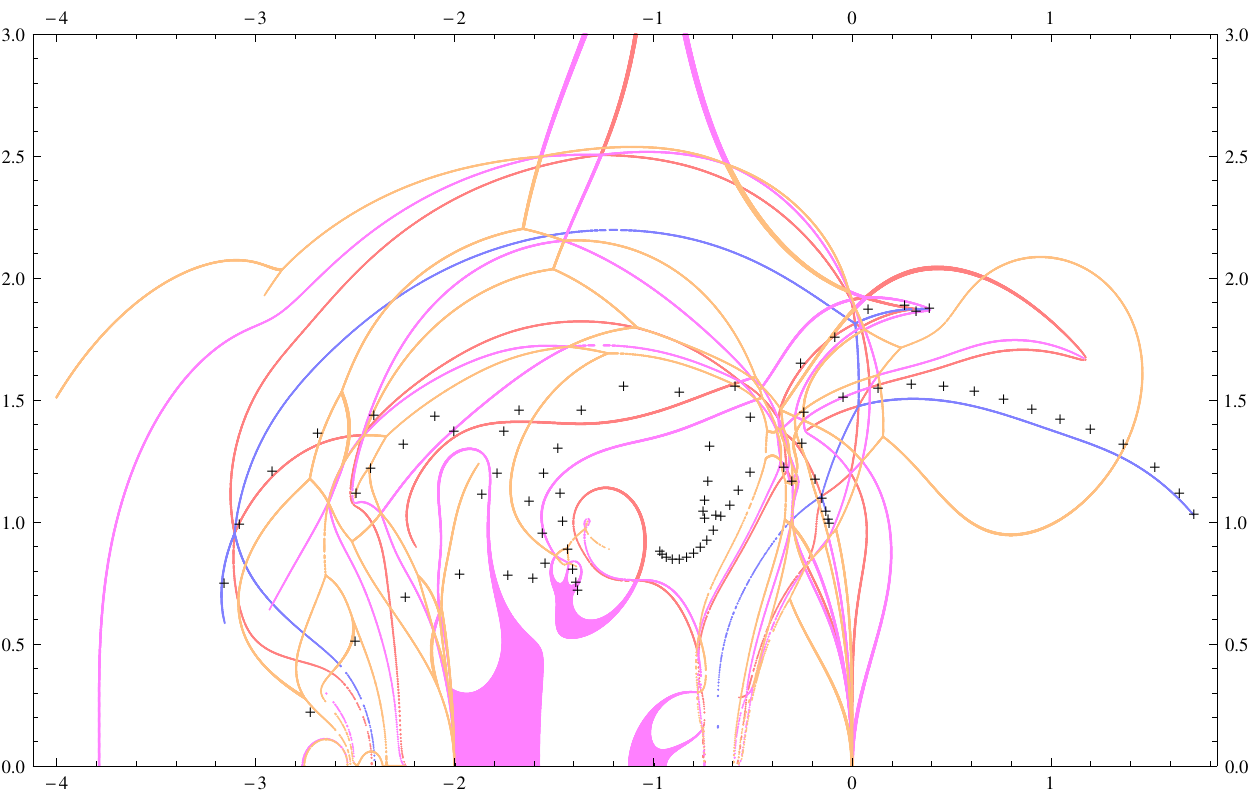}~~~~~
\includegraphics[width=0.47\textwidth]{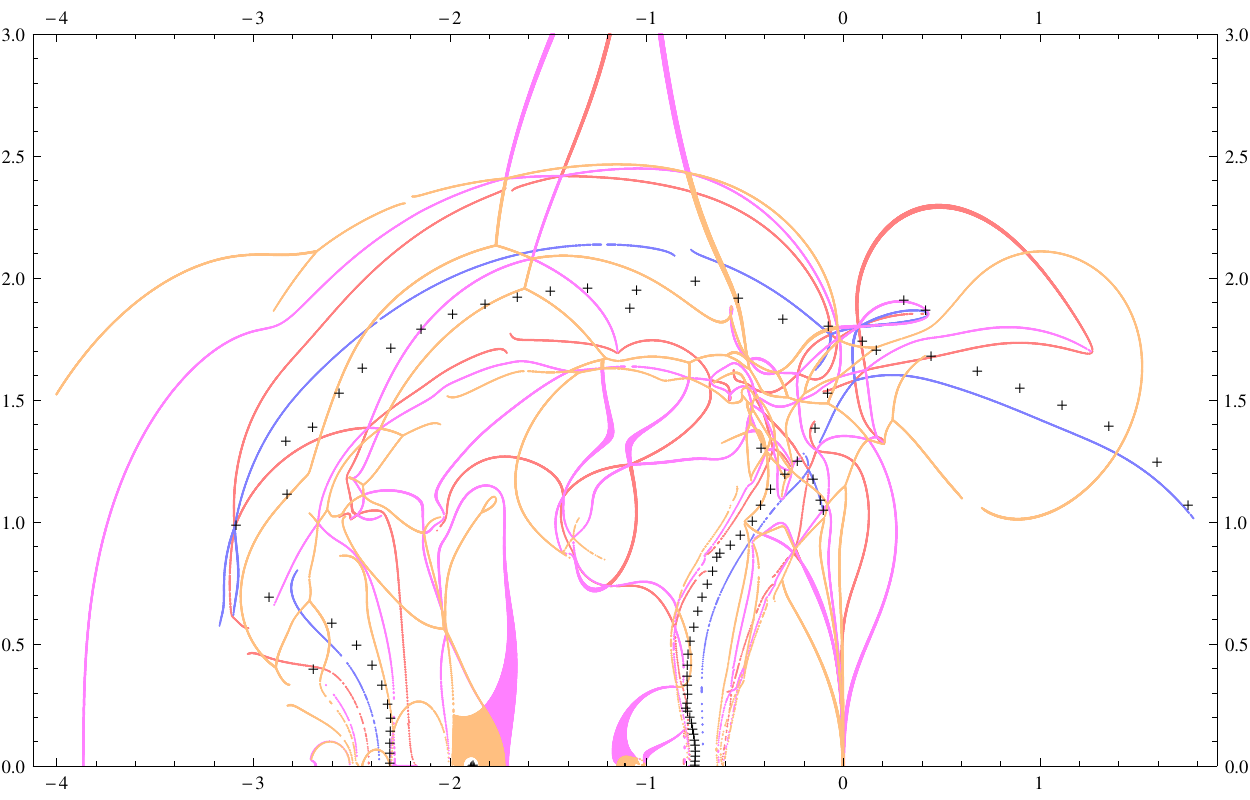}
\end{center}
\hspace{2.5 cm}(e) $m=4$, $q=2$, $n=20$, \hspace{4.5 cm}(f) $m=4$, $q=2.1$, $n=10$.
\caption{Diagrams of eigenvalue-crossings and Fisher zeros in the complex $v$ plane for strips of the kagome lattice of width $m$. The crosses correspond to the exact zeros for strips of length $n$. Clearly for the case $q=1$ all the zeros lie at $v=-1$ for any length $n$. The lines correspond to the geometric zones where is satisfied $f_{ij}\equiv|\lambda_i|-|\lambda_j|=0$ (within a numerical tolerance given by $||\lambda_i|-|\lambda_j||/\sqrt{|\lambda_i|^2+|\lambda_j|^2}\le 10^{-3}$), where the colorations means: blue for $f_{12}$, red for $f_{23}$, magenta for $f_{34}$ and orange for $f_{45}$. The others combinations, giving either empty sets or minor regions, are not displayed here.} \label{Fisher_kago}
\end{figure}

Complementary to the thermodynamic functions, we drawn the diagrams for eigenvalues crossings and the exact Fisher zeros for strips of finite length (Fig. \ref{Fisher_kago}). The crossings are given by the equations $f_{ij}\equiv|\lambda_i|-|\lambda_j|=0$. The diagrams show interesting examples of degeneration in the context of the Beraha-Kahane-Weiss theorem, and of crossing of three simple eigenvalues, see \cite{Beraha_79,Beraha_80} and T-points in \cite{Salas01}. To understand this we have considered the case of $m=3$ or 4 where $N=5$ or 14 respectively. From the simplest case $m=3$, $q=1$ (Fig. \ref{Fisher_kago} (a)) can be readily seen that the curve $f_{12}=0$ is not faithful to the locus of the zeros, which is for this case simply a point located at $v=-1$. This is consequence of a degeneration of some of the coefficients $\alpha_1$ and/or $\alpha_2$ over the curve $f_{12}=0$. This degeneration can be understood as a consequence of a higher symmetry expected to appear at integer-valued numbers of $q$, corresponding to a higher degeneration of the energy levels for instance. To verify this affirmation the diagram with a {\it perturbed} non-physical value of $q=1.1$ is considered as well (Fig. \ref{Fisher_kago} (b)), where the crosses spread out next to the line $f_{12}=0$. This fact repeats for $q=2$, where in the antiferromagnetic regime the Fisher zeros tend to be located next to the curve given by $f_{23}=0$ rather to the one given by $f_{12}=0$. Going on higher values of $q$ is more unlikely have this kind of degeneracy. It is worth to mention that this considerations are very relevant in order to properly take the thermodynamic limit. Actually it can be noted the line $f_{12}=0$ cross the physical values of the temperature in the antiferromagnetic regime for unphysical values of $q\in(q_1,q_2)- \{1,2\}$ where $q_1<1$ and $q_2>2$. The knowledge of the limiting values of $q_1$ and $q_2$ requires a further study. It would be interesting to know if this phenomena could be related to the appearance of a spontaneous magnetization when a external magnetic field is applied.

\section{Further examples}

In previous sections we have studied periodic strip lattices. Nevertheless the main idea, namely to derive a homogeneous first order linear recurrence relation for a vector whose the first component is the partition function itself, can be applied even in a slightly generalized cases. As illustrations we consider a lattice in which every layer have a composite structure, and a lattice in which the layer varies along the longitudinal direction of the lattice. In this sense, the present formalism is an extension of the classic results of Biggs, et. al. \cite{Biggs_72} and of Beraha, et. al. \cite{Beraha_79,Beraha_80} in which the authors considered a homogeneous higher order linear recurrence relation with ``constant coefficients"\footnote{Namely, a linear recurrence relation whose coefficients do not depend explicitly on the recursive step $n$.} for a scalar in order to find the partition function of periodic strip lattices.

\subsection{Ornamented $L_y=3$ hybrid strip}\label{ornamented}

\begin{figure}[t]
\begin{center}
\hspace{1cm} \includegraphics[width=0.14\textwidth]{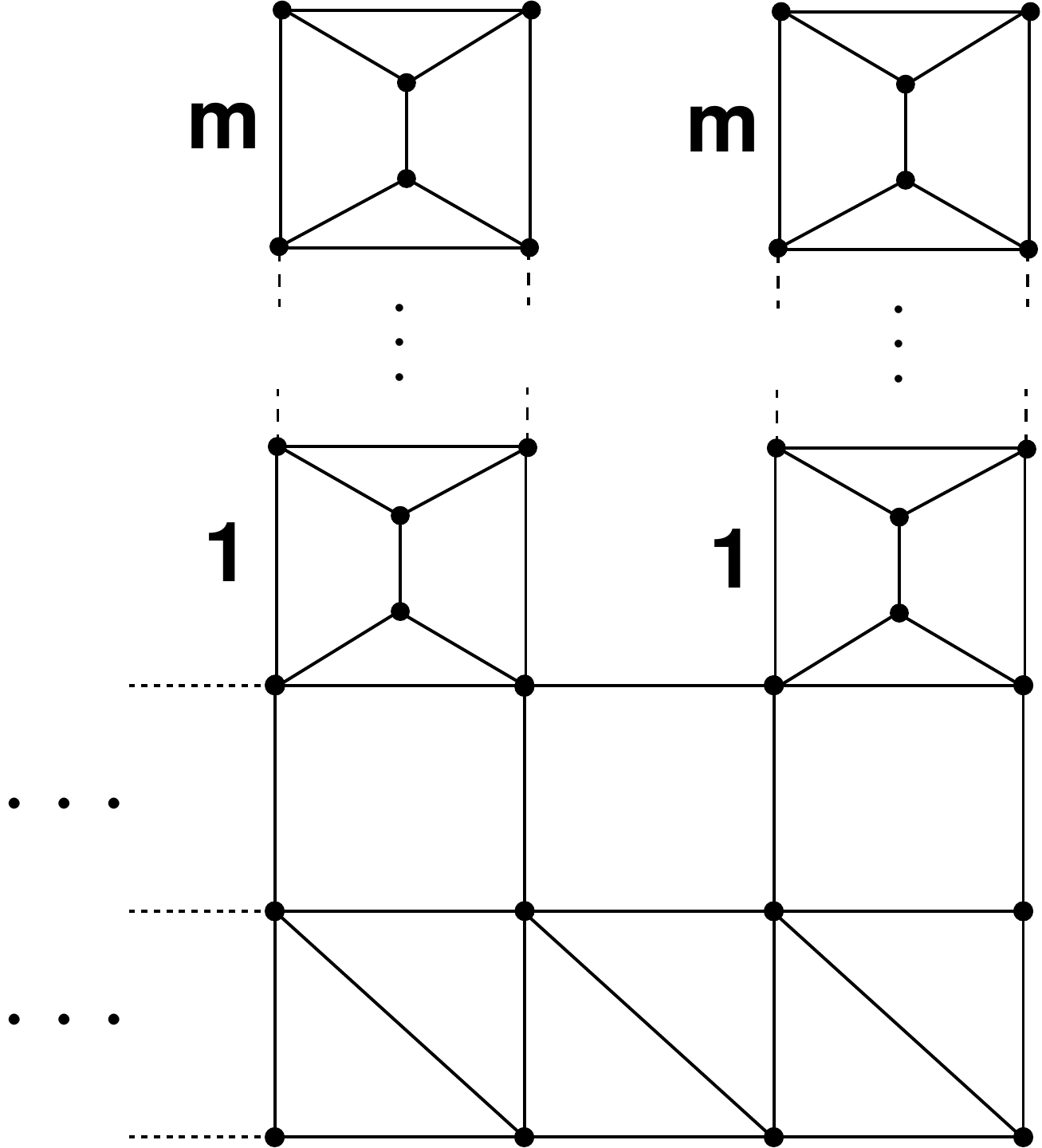}  \hspace{1.8cm} \includegraphics[width=0.14\textwidth]{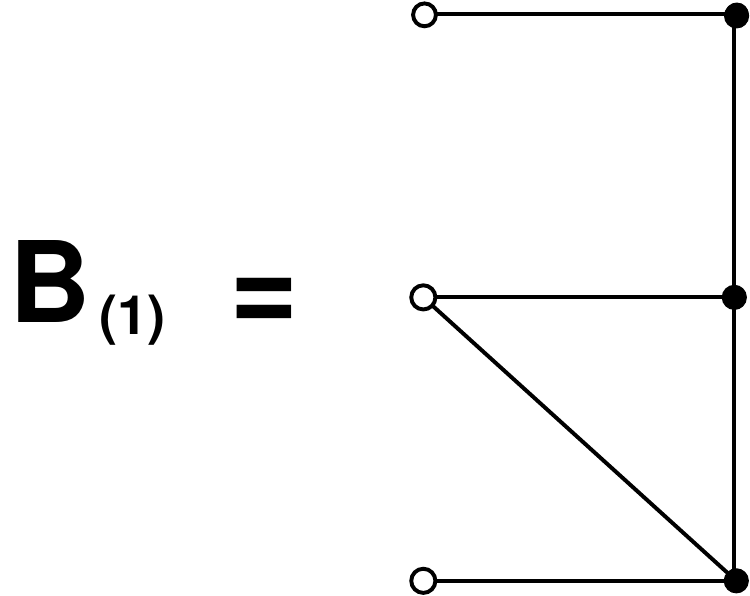} \hspace{0.5cm} \includegraphics[width=0.14\textwidth]{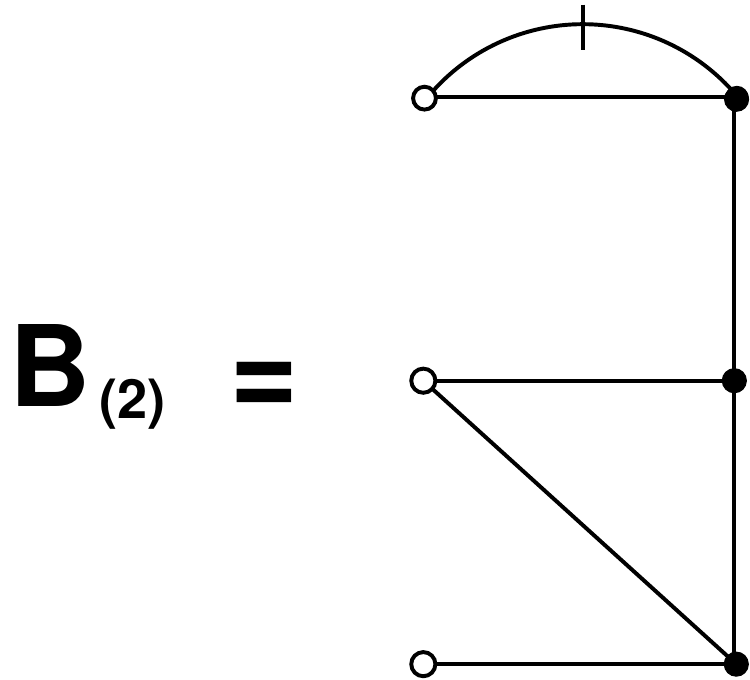} \\
(a) \hspace{6cm} (b) \hspace{12cm}
\end{center}
\caption{(a) Right end of a mixture of triangular and square lattices ``ornamented'' periodically by a series of shortest-path strips of length $m$. (b) Intermediate blocks $B_{(1)}$ and $B_{(2)}$ to be considered in the calculation of the full transfer matrix of the structure in (a).}
\label{complex}
\end{figure}

The first example correspond to a composite periodic structure depicted in Fig. \ref{complex}. It consists of a mixture of square and triangular strips decorated by segments of the shortest-path lattice. Obviously, one expects that for such a lattice the construction of the exact solution should become much more difficult. The reasons are, firstly, that the periodicity is not at every layer, and secondly that the lattice has not one but three different types of basic boxes (square, triangular and shortest path in the above example). Thus, the system has less symmetries and the dimension of the corresponding transfer matrix increases.

\begin{figure}[t]
\begin{center}
\includegraphics[width=0.4\textwidth]{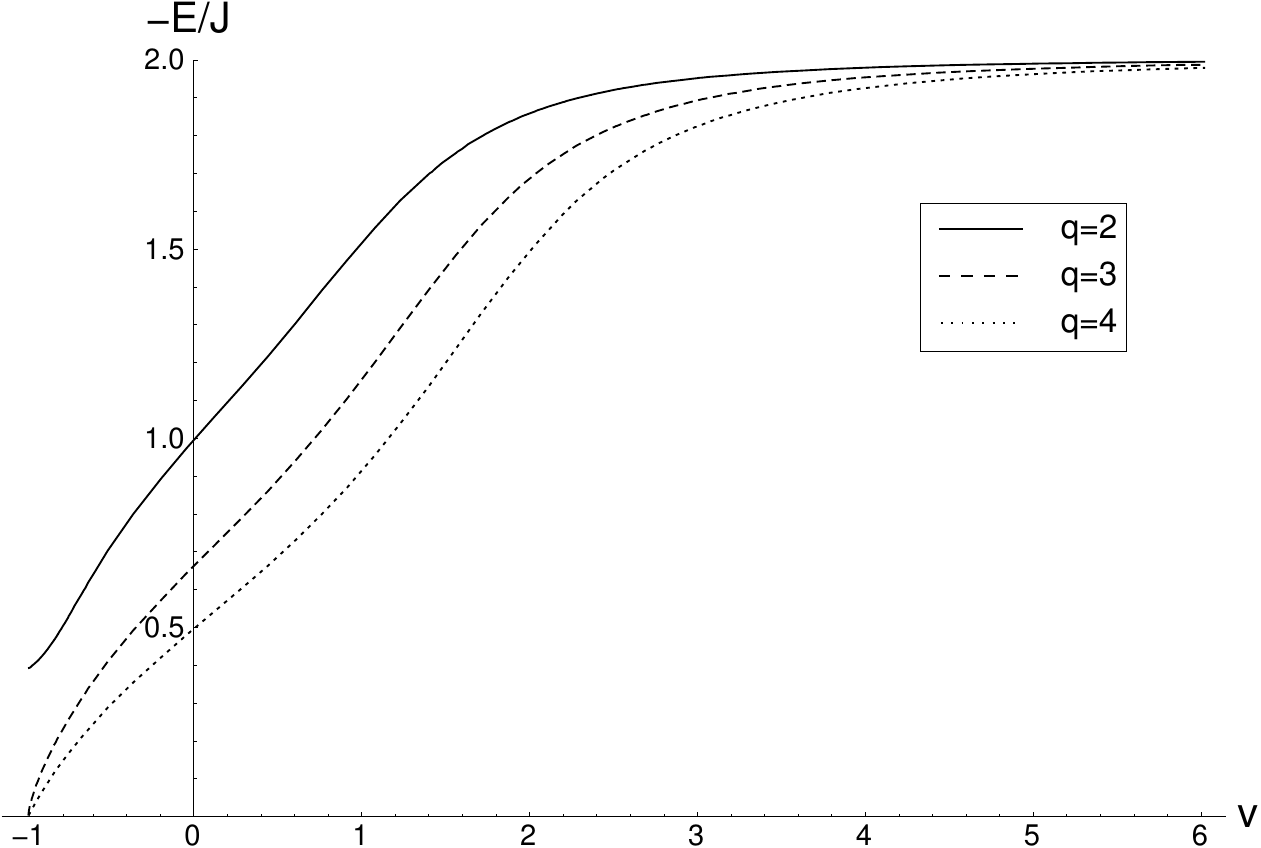}~~~~~~~~~
\includegraphics[width=0.4\textwidth]{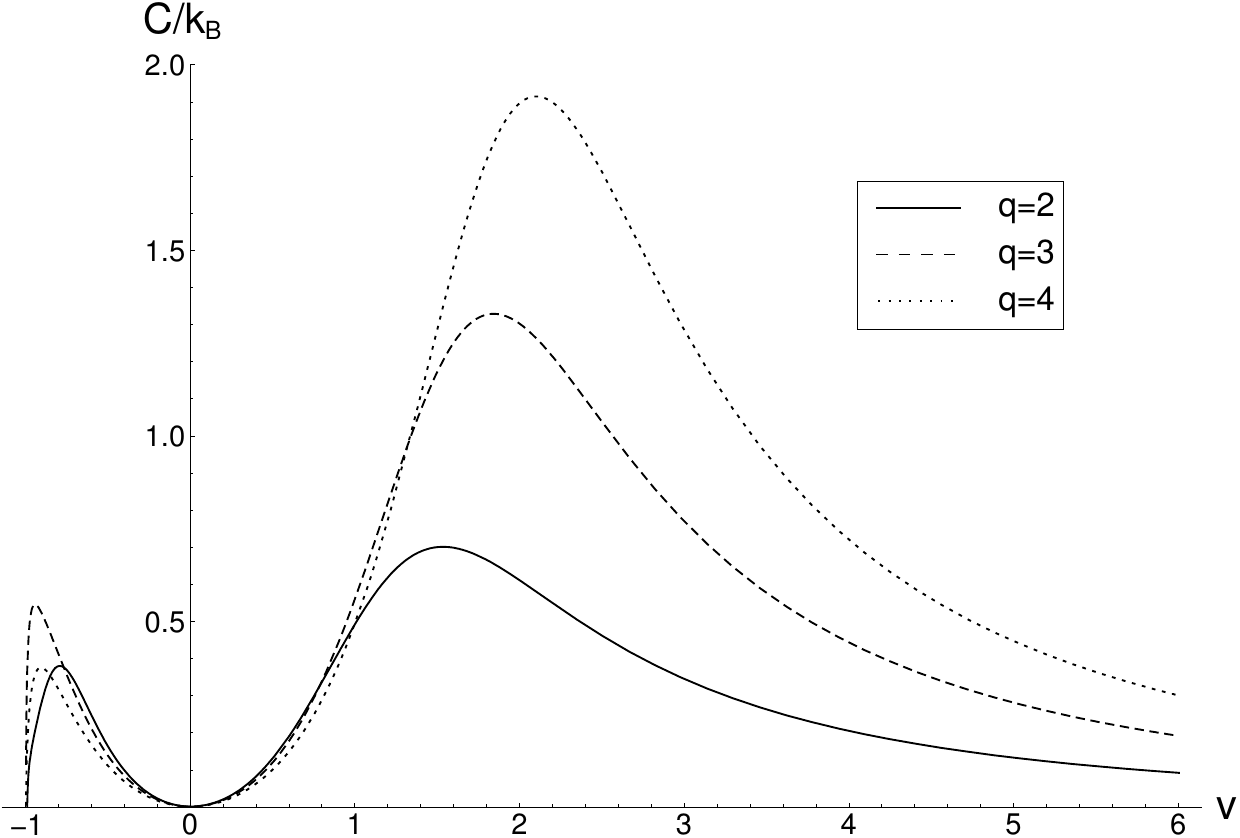}\\[0pt]
(a) \hspace{9cm} (b)
\end{center}
\caption{Results for the ``ornamented" lattice strip in Fig. \ref{complex}a. (a) Reduced internal energy as a function of the temperature parameter $v=e^K-1$ for different values of $q$. (b) Dimensionless specific heat as a function of $v$.}
\label{strange_plots}
\end{figure}

To deal with this lattice it is convenient to think of the problem in different steps. First, we decompose the ornament of the last block making use of the corresponding transfer matrix calculated in this case in Sec. \ref{sh_path}. It is easy to notice that at the end of this process we end up with two types of blocks shown in Fig. \ref{complex}(b). Following again the method described in Sec. \ref{method}, we can then calculate their corresponding $5\times 5$ transfer matrices $A^{(1)}$ and $A^{(2)}$ that enable us to decompose the main body of the strip. Lastly, by combining these steps we obtain,
\begin{equation}\label{ornament_sol}
\overrightarrow{Z}(n)=\left(\alpha(m) A^{(1)} + \beta(m) A^{(2)}\right)A^{(1)}  \overrightarrow{Z}(n-1),
\end{equation}
where $\alpha(m)$ and $\beta(m)$ are scalar quantities given by
\begin{equation}
\alpha(m)=\left\{\left(\mathbf{A}^{\text{sh}}(q,v)\right)^{m}\right\}_{1,1}, \qquad \beta(m)=\left\{\left(\mathbf{A}^{\text{sh}}(q,v)\right)^{m}\right\}_{1,2}.
\end{equation}
The details of the calculation and explicit expressions for all of the matrices involved can be found in the Appendix. Using the same method as in previous cases, we can extract the internal energy and specific heat displayed in Fig. \ref{strange_plots}.

Finally, we point out that the solution outlined here can be straightforwardly generalized for any type of $L_y=2$ ornament for which the transfer matrix is known. One only needs to replace $\alpha(m)$ and $\beta(m)$ accordingly in (\ref{ornament_sol}).

\subsection{Step-dependent transfer matrix}\label{step-dependent}

Up to now, we have only analyzed cases in which the lattice has a periodic recursive symmetry, namely always there is a layer (composite or not) that is periodically repeated. However the key idea of the present method also works for recursive non-periodic lattices.

\begin{figure}[t]
\begin{center}
\includegraphics[width=0.2\textwidth]{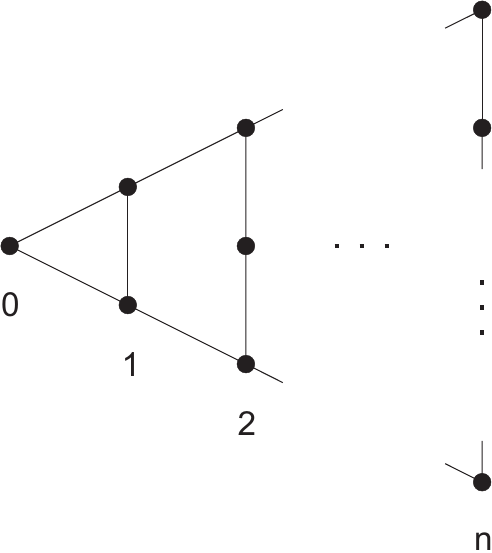}
\end{center}
\caption{Example of a lattice in which the transfer matrix of every layer explicitly depends on the recursive step $n$.}
\label{n-dep}
\end{figure}

As a simple example let us consider the lattice depicted in the Fig. \ref{n-dep}. The resulting linear system will indeed have an $n$-dependent transfer matrix\footnote{It is worth to note that in the present example each layer can be contracted to just a single edge, by using the usual "series reduction" principle, however there is possible to consider more complicated cases in which that reduction procedure in no longer applicable.},
\begin{equation}
\overrightarrow{Z}(n)=A(n)\overrightarrow{Z}(n-1)
\end{equation}
with,
\begin{equation}
\label{matrixn}
A(n)=\left(
          \begin{array}{cc}
            \frac{1}{q}(T(n+2)-q v^{n+2}) & v^{n+2} \\
            \frac{q+2v}{q} T(n) & \frac{v^2}{q} T(n) \\
          \end{array}
        \right),
\end{equation}
where $T(n)$ stands for the partition function of the ring with $n$ vertices given by
\begin{equation}
T(n)=q v^n +q \sum_{i=0}^{n-1} v^i (q+v)^{n-1-i}.
\end{equation}
The $n$-dependence in (\ref{matrixn}) is a manifestation of the fact that each new layer in the lattice is not the same as the previous one. Nevertheless, it is possible to apply the method in exactly the same way as before. The vector of the partition function can be written in the form
\begin{equation}
\label{sol_n-dep}
\overrightarrow{Z}(n)=A(n) A(n-1)... A(1) \overrightarrow{Z}(0),
\end{equation}
where the boundary term is given by the two component vector $\overrightarrow{Z}(0)=(q,q)$. The partition function can be read from the first component of (\ref{sol_n-dep}). Moreover, it turns to be quite complicated to solve in general the characteristic equation corresponding to the recursive equation of the present example because the coefficients depends on $n$.

\begin{figure}[t]
\begin{center}
\includegraphics[width=0.2\textwidth]{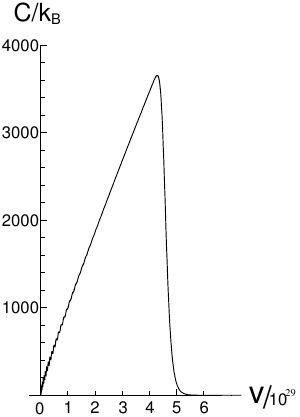}~~~~~~~~~~~~
\includegraphics[width=0.45\textwidth]{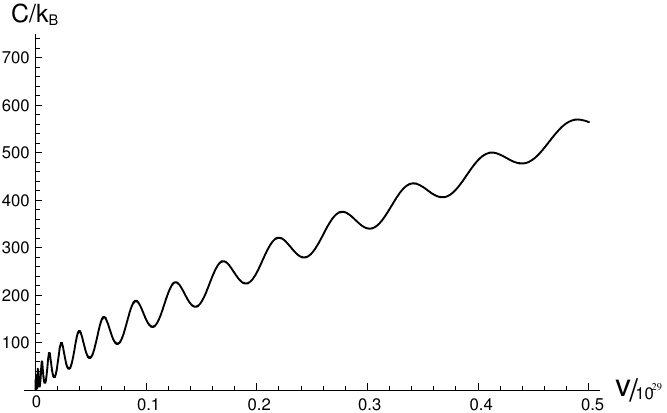}~~~~~~~~~~~~\\
(a)\hspace{5.8cm}(b)\hspace{3cm}
\end{center}
\caption{(a) Specific heat in the ferromagnetic range for the $n=50$ lattice. A global maximum and many local peaks in the high temperature regime are observed. (b) Lower temperature regime in more detail. The density of local peaks is observed to increase with temperature.}
\label{n-dep_spec_heat}
\end{figure}

The specific heat of this lattice shows an interesting behavior presenting many peaks which are consequences of peaks of the specific heat of all the $T(n')$ ($n'=1,...,n$) sublattices existing in $Z(n)$, see Fig. \ref{n-dep_spec_heat}.

\section{Conclusions and perspectives}\label{conc}

We have exploited a method based on the deletion-contraction theorem to obtain a variety of new exact results for the partition function of the Potts model. This technique has been applied in many cases in which the lattice geometry differs from the squared one. Indeed, the only requirement for the present method apply is that the lattice has a recursive layered structure, while no extra symmetries are needed.

In the case of strips lattices, we have shown that the technique is {\it not} restricted to the simpler cases of $L_y=2$ or $3$. With the aid of a computer program we were able to find the exact transfer matrix for the $L_y=5$ kagome strip. In a future work we intend to use the same approach to obtain exact results for the $L_y=5$ diced lattice strip.

It has also been shown that this procedure can be easily applied to layered lattices whose widths changes periodically, or change in a pre-defined way. The possibility to analyze other lattices is worth to be consider and could have important applications.

It is worth noting that this technique can be extended to deal with other boundary conditions such as periodic, toroidal or M\"{o}bius. Wider strips and non planar strips can be considered as well.

\section*{Appendix}
As explained in the main text, one can proceed by first decomposing the ornament and then the main body of the strip. Combining both of these procedures the resulting transfer matrix is given by,
\begin{equation}
\overrightarrow{Z}(n)=\left(\alpha(m) A^{(1)}+\beta(m) A^{(2)}\right) A^{(1)}  \overrightarrow{Z}(n-1),
\end{equation}
where $\alpha(m)$ and $\beta(m)$ are scalar quantities that can be extracted directly from the transfer matrix of the ``ornamented'' strip,
\begin{equation}
\alpha(m)=\left\{\left(\mathbf{A}^{\text{sh}}(q,v)\right)^{m-1}\right\}_{1,1}, \qquad \beta(m)=\left\{\left(\mathbf{A}^{\text{sh}}(q,v)\right)^{m-1}\right\}_{1,2}.
\end{equation}
$A^{(1)}$ and $A^{(2)}$ are $5\times5$ matrices given by
\begin{equation}
A^{(1)}=\left(
          \begin{array}{ccccc}
            B_{s_1,s_2}^{c_1,c_2,c_3} & C_{s_2}^{c_1,c_3} & D_{s_1,s_2}^{c_2,c_4,c_5} & E_{s_2}^{c_2,c_4,c_5} & F_{s_2}^{c_2} \vspace{0.1cm}\\
            B_{t_1,t_2}^{c_1,c_2,c_3} & C_{t_2}^{c_1,c_3} & D_{t_1,t_2}^{c_2,c_4,c_5} & E_{t_2}^{c_2,c_4,c_5} & F_{t_2}^{c_2} \vspace{0.1cm}\\
            B_{s_1,s_2}^{d_1,d_2,d_3} & C_{s_2}^{d_1,d_3} & D_{s_1,s_2}^{d_2,d_4,d_5} & E_{s_2}^{d_2,d_4,d_5} & F_{s_2}^{d_2} \vspace{0.1cm}\\
            B_{t_1,t_2}^{d_1,d_2,d_3} & C_{t_2}^{d_1,d_3} & D_{t_1,t_2}^{d_2,d_4,d_5} & E_{t_2}^{d_2,d_4,d_5} & F_{t_2}^{d_2} \vspace{0.1cm}\\
            \tilde{B}_{c_1,c_2,c_3;d_1,d_2,d_3}^{e_1,e_2,e_3,e_4} & \tilde{C}_{e_3,e_4} & \tilde{D}_{c_2,c_4,c_5;d_2,d_4,d_5}^{e_1,e_2} & \tilde{E}_{e_3,e_4} & \tilde{F}_{e_3,e_4}\\
          \end{array}
        \right),
\end{equation}
\begin{equation}
A^{(2)}=(1+v)\left(
          \begin{array}{ccccc}
            B_{s_3,s_4}^{c_1,c_2,c_3} & C_{s_4}^{c_1,c_3} & D_{s_3,s_4}^{c_2,c_4,c_5} & E_{s_4}^{c_2,c_4,c_5} & F_{s_4}^{c_2} \vspace{0.1cm}\\
            B_{t_3,t_4}^{c_1,c_2,c_3} & C_{t_4}^{c_1,c_3} & D_{t_3,t_4}^{c_2,c_4,c_5} & E_{t_4}^{c_2,c_4,c_5} & F_{t_4}^{c_2} \vspace{0.1cm}\\
            B_{s_3,s_4}^{d_1,d_2,d_3} & C_{s_4}^{d_1,d_3} & D_{s_3,s_4}^{d_2,d_4,d_5} & E_{s_4}^{d_2,d_4,d_5} & F_{s_4}^{d_2} \vspace{0.1cm}\\
            B_{t_3,t_4}^{d_1,d_2,d_3} & C_{t_4}^{d_1,d_3} & D_{t_3,t_4}^{d_2,d_4,d_5} & E_{t_4}^{d_2,d_4,d_5} & F_{t_4}^{d_2} \vspace{0.1cm}\\
            \tilde{B}_{c_1,c_2,c_3,d_1,d_2,d_3}^{f_1,f_2,f_3,f_4} & \tilde{C}_{f_3,f_4} & \tilde{D}_{c_2,c_4,c_5,d_2,d_4,d_5}^{f_1,f_2} & \tilde{E}_{f_3,f_4} & \tilde{F}_{f_3,f_4}\\
          \end{array}
        \right),
\end{equation}
where,
\begin{eqnarray}
B_{s_1,s_2}^{c_1,c_2,c_3}&=&s_1((q+v) c_1+c_2+ (1+v) c_3) +s_2 c_1,\\
C_{s_2}^{c_1,c_3}&=&s_2(v c_1+ (1+v) c_3),\\
D_{s_1,s_2}^{c_2,c_4,c_5}&=&s_1(v c_2+(q+v) c_4 +(1+v) c_5) +s_2 c_4,\\
E_{s_2}^{c_2,c_4,c_5}&=&s_2(v c_2+v c_4+(1+v) c_5),\\
F_{s_2}^{c_2}&=& s_2 c_2,\\
\tilde{B}_{c_1,c_2,c_3,d_1,d_2,d_3}^{e_1,e_2,e_3,e_4}&=& e_1 ((q+v)c_1 +c_2 +(1+v) c_3)+ e_2 ((q+v) d_1 +d_2+(1+v) d_3) \\ \nonumber
                                                     &~&      +(q+2v) e_3 +(1+v) e_4,\\
\tilde{C}_{e_3,e_4}&=& v (v + T_2/q) e_3 + v(1+v)(2+v) e_4,\\
\tilde{D}_{c_2,c_4,c_5,d_2,d_4,d_5}^{e_1,e_2}&=&e_1 (v c_2 +(q+v) c_4 +(1+v) c_5) + e_2 (v d_2 +(q+v) d_4 +(1+v) d_5),\\
\tilde{E}_{e_3,e_4}&=& v^2 (v + T_2/q) e_3 + v^2(1+v)(2+v) e_4,\\
\tilde{F}_{e_3,e_4}&=&v (q+2v) e_3 + v (1+v) e_4,\\
\end{eqnarray}
with $T_2 = q (q+2v) + q v^2$, and
\begin{eqnarray}
&s_1=q+2v, \qquad s_2=v^2, \qquad s_3=1, \qquad s_4=v,&\\
&t_1=1+v, \qquad t_2=v(1+v), \qquad t_3=0, \qquad t_4=1+v,&\\
&c_1=q+3v, \qquad c_2=v^2, \qquad c_3=v^2, \qquad c_4=v^2, \qquad c_5=v^3,&\\
&d_1=(1+v), \qquad d_2=v(1+v), \qquad d_3=v(1+v), \qquad d_4=0, \qquad d_5=v^2 (1+v),&\\
&e_1 = 1, \qquad e_2 = v, \qquad e_3 = v, \qquad e_4 = v^2,&\\
&f_1 = 0, \qquad f_2 = 0, \qquad f_3 = 1, \qquad f_4 = v.&\\
\end{eqnarray}
As usual the general expression for the vector of partition functions is given by
\begin{equation}
\overrightarrow{Z}(n)=\left[\left(\alpha(m) A^{(1)}+\beta(m) A^{(2)}\right) A^{(1)}\right]^{n-1}  \overrightarrow{Z}(1),
\end{equation}
where the initial data correspond to the vector
\begin{equation}
\overrightarrow{Z}(1)=\left(
                        \begin{array}{c}
                          q(q+v)^2 \\
                          q(q+v)(1+v) \\
                          q(q+v)(1+v) \\
                          q(1+v)^2 \\
                          q(q+v)+q v(1+v) \\
                        \end{array}
                      \right)
\end{equation}

\textbf{Acknowledgements}\\
{\small {The authors wish to give a very warm thank for his illuminating comments and suggestions to Marco Astorino who participated to the early stage of this project. The authors also acknowledge invaluable help from Marcos P\'erez. S. A. R. wishes to thank Rafael Benguria for kind support and many fruitful discussions. This work was supported by FONDECYT grants 11080056, 3100140, 11110537; Anillos de Investigaci\'on en Ciencia y Tecnolog\'{\i}a, projects ACT-56, Lattice and Symmetry, and by the Southern Theoretical Physics Laboratory ACT- 91 grants from CONICYT. S. A. R. acknowledges financial support from VRAID (PUC) and Facultad de Fisica (PUC). The Centro de Estudios Cient\'{\i}ficos (CECs) is funded by the Chilean Government through the Centers of Excellence Base Financing Program of CONICYT.}}

\end{document}